\documentclass[pre,aps,twocolumn,amsmath,amssymb,amsfonts,floatfix,superscriptaddress,longbibliography]{revtex4-1} 
\usepackage{graphicx}
\usepackage{subfigure} 
 \usepackage{color} 
 
 \definecolor{vinous}{rgb}{0.74,0.08,0.12}
 
 \definecolor{IndianRed}{rgb}{0.804,0.361,0.361}

 \newcommand{\red}[1]{\textcolor{black}{#1}}

\begin{document} 
 
\title{Binding potential and wetting behaviour of binary liquid mixtures on surfaces} 
 
\author{Mounirah Areshi} 
\affiliation{Department of Mathematical Sciences, Loughborough University, Loughborough LE11 3TU, United Kingdom} 
\affiliation{Department of Mathematics, Faculty of Science, University of Tabuk, P.O. Box 741, Tabuk 71491, Saudi Arabia} 

\author{Dmitri Tseluiko}
\affiliation{Department of Mathematical Sciences, Loughborough University, Loughborough LE11 3TU, United Kingdom}
\affiliation{Interdisciplinary Centre for Mathematical Modelling, Loughborough University, Loughborough LE11 3TU, United Kingdom} 
 
\author{Uwe Thiele}
\email{u.thiele@uni-muenster.de}
\homepage{http://www.uwethiele.de}
\thanks{ORCID ID: 0000-0001-7989-9271}
\affiliation{Institute of Theoretical Physics, University of M\"unster, Wilhelm-Klemm-Str.\ 9, 48149 M\"unster, Germany}
\affiliation{Center for Nonlinear Science (CeNoS), University of M\"unster, Corrensstr.\ 2, 48149 M\"unster, Germany}

\author{Benjamin D. Goddard}
\affiliation{School of Mathematics and the Maxwell Institute for Mathematical Sciences, University of Edinburgh, Edinburgh EH9 3FD, UK}

\author{Andrew J. Archer} 
\email{A.J.Archer@lboro.ac.uk} 
\affiliation{Department of Mathematical Sciences, Loughborough University, Loughborough LE11 3TU, United Kingdom}
\affiliation{Interdisciplinary Centre for Mathematical Modelling, Loughborough University, Loughborough LE11 3TU, United Kingdom}
 
\begin{abstract} 
We present a theory for the interfacial wetting phase behaviour of binary liquid mixtures on rigid solid substrates, applicable to both miscible and immiscible mixtures. In particular, we calculate the binding potential as a function of the adsorptions, i.e.\ the excess amounts of each of the two liquids at the substrate. The binding potential fully describes the corresponding interfacial thermodynamics. Our approach is based on classical density functional theory. Binary liquid mixtures can exhibit complex bulk phase behaviour, including both liquid-liquid and vapour-liquid phase separation, depending on the nature of the interactions between all the particles of the two different liquids, the temperature and the chemical potentials. Here we show that the interplay between the bulk phase behaviour of the mixture and the properties of the interactions with the substrate gives rise to a wide variety of interfacial phase behaviours, including mixing and demixing situations. We find situations where the final state is a coexistence of up to three different phases. We determine how the liquid density profiles close to the substrate change as the interaction parameters are varied and how these determine the form of the binding potential, which in certain cases can be a multi-valued function of the adsorptions.  We also present  profiles for sessile droplets of both miscible and immiscible binary liquids.
\end{abstract} 
 \maketitle 
  \section{Introduction}\label{sec:1}
%
The behaviour of liquids and liquid mixtures on rigid solid substrates with planar surfaces is of great relevance to many areas of life, such as in food processing and preparation, oil recovery, cosmetics, pharmaceutical and a host of other industries. The interfacial phase behaviour of even simple one-component liquids at such walls can be varied and complex. Typically, at low temperatures liquids only partially wet solids. However, for most substrates, as the temperature $T$ is increased, the contact angle $\theta$ of sessile droplets decreases until at some temperature $T_w < T_c$ there is a wetting transition, where $T_w$ is the wetting temperature and $T_c$ is the bulk fluid critical temperature \cite{thinfilms, dietrich88, evans2019unified}. Additionally, there is the possibility of a drying transition occurring at substrates which very weakly interact with the liquid \cite{evans2019unified}. The nature of the wetting behaviour all depends, amongst other properties of the system, on the form and strength of the interparticle and particle-wall interactions. When the liquid at the interface is a mixture of two different species, then the interfacial phase behaviour becomes even more complicated. In the same way that the interfacial phase behaviour of a one-component system is connected to the bulk phase behaviour (i.e.\ interfaces can promote incipient phases when the system is near to bulk phase coexistence), likewise, the interfacial phase behaviour of binary mixtures is related to and can influence the bulk phase behaviour \cite{GuBR1991pa, BiNP1997zpbm, FiMD1997prl, KEMR2001cpc, ThMF2007pf, madruga2009decomposition, bribesh2012decomposition, SHY2023s}. This can involve vapour-liquid and/or liquid-liquid phase separation, or even an interplay of both. In particular, thin films of mixtures on solid substrates, e.g.\ polymer blends or molten alloys, may undergo combined dewetting and decomposition processes \cite{BrBr1992prl, COFC2007m, OKNM2008p, ThTL2013prl, DGGR2021l}. Here, the behaviour depends on the nature of the interaction potentials between the different species of particles, the temperature and the chemical potentials and also on the properties of the interactions between the liquid particles and the wall. For example, if the wall favours one species over the other, then this can induce various types of ordering in a film at a wall \cite{madruga2009decomposition, bribesh2012decomposition, malijevsky2013sedimentation}.

A key quantity that characterises \red{interfacial} wetting behaviour is the binding potential $g$\red{. This incorporates} the contribution to the free energy of the system due to the solid-liquid interface influencing the vapour-liquid interface\red{. This occurs} either when these two interfaces are close to one another, or if they merge so that there is just a solid-vapour interface \cite{dietrich88, schick1990introduction, bonn01, de2013capillarity}. For a one component liquid, $g$ can be written as a function of the thickness of the liquid film on the surface $h$, i.e.\ $g = g(h)$. However, since the \red{density distribution profiles vary smoothly between the liquid and gas phases}, $h$ is ill-defined (particularly when it is small), and it is arguably better to consider the binding potential as a function of the adsorption $\Gamma$ (see Eq.~\eqref{eq:adsorption_alpha} below), i.e.\ $g = g(\Gamma)$ \cite{hughes2015liquid}. Such binding potentials (or wetting energies), once extracted from microscopic models \cite{tretyakov2013, MBKP2014acis, hughes2017, buller2017nudged} or appropriate approximations \cite{DeLa1941apu, DzLP1960spj, Isra1972prsla}, are employed as crucial elements of mesoscale hydrodynamic models where they enter as the Derjaguin (or disjoining) pressure \cite{de2013capillarity, craster2009dynamics, bonn2009wetting}.

In a similar manner, for binary liquid mixtures composed of two different species of particles, which we refer to here as species-$A$ and species-$B$, one can define the binding potential $g(\Gamma_{A}, \Gamma_{B})$, as a function of the adsorptions of the two different species at the interface, $\Gamma_{A}$ and $\Gamma_{B}$. Of course, if the liquids are immiscible and there is a film of one liquid on top of the other liquid at the surface, then one could equally consider the binding potential to be a function of the thicknesses of the two films, $h_A$ and $h_B$. However, if they are miscible liquids, then considering the binding potential to be a function of the adsorptions $\Gamma_{A}$ and $\Gamma_{B}$ is more meaningful. Such binding potentials have been postulated previously, employing simple dependencies for concentration-dependent wetting energies \cite{ThTL2013prl, ThAP2016prf}. What is clear e.g.\ from the results in Refs.~\cite{DeCh1977jcis, Boud1987jp, FoBr1997el, WoSc2004jpm, FoBr1998m, FoBr1999pa} is that the behaviour of mixtures at interfaces can be very rich.

The adsorptions are defined as the excess amount of each species at the wall:
\begin{equation}
\label{eq:adsorption_alpha}
\Gamma_{\alpha} = \int_0^{\infty} \left (\rho_{\alpha}(z) - \rho_{\alpha}^{b}\right) dz,
\end{equation} 
where $\rho_{\alpha}(z)$ and $\rho_{\alpha}^{b}$ are the density profile and the bulk (vapour) density of species $\alpha$, respectively, ($\alpha=A$ or $B$) and $z$ is the direction perpendicular to the wall. In our calculations below, to simplify the description of the binary mixture, we map the particle densities onto a lattice model, with dimensionless density for each species $\rho_{\mathbf{i}}^{\alpha}$ at each lattice site $\mathbf{i}$, i.e.\ one can consider the lattice densities to be defined as:
\begin{equation}
\rho_{\mathbf{i}}^{\alpha} = \int_{{{vol}\;\mathbf{i}}} {\rho_\alpha({\mathbf{r}}}) d\mathbf{r},  
\end{equation} 
where the integral is over the volume of lattice site $\mathbf{i}$.

Our approach here for calculating the binding potential is based on the lattice density functional theory (DFT) approach of Hughes et al.\ \cite{hughes2015liquid}. This, in turn, is based on classical DFT \cite{evans1979nature, hansen2013theory}, and uses a constrained Picard iteration minimisation approach \cite{hughes2014introduction, hughes2015liquid} to solve the DFT equations for mixtures subject to the constraints that the adsorptions of the different species are the specified values. This approach was first developed in the context of studying nucleation \cite{archer2011nucleation}, where it was shown that these constraints are equivalent to applying an additional external field to stabilise the specified amount of liquid at the interface -- see also \cite{hughes2015liquid, hughes2017, buller2017nudged}. Here, we extend and apply the approach in order to study liquid mixtures at solid substrates.

The method is valid over the full temperature range where liquid-vapour coexistence occurs and the whole range of concentrations of the two liquids. We are able to determine the excess free energy as a function of the excess amount of each phase adsorbed at the wall. Moreover, we determine the form of the density profiles of each of the species, showing how the phase-separation of the mixture (if it occurs) and the properties of the interactions with the wall and between species influences the form of the density profiles \cite{evans1979nature, hansen2013theory, llovell2010classical}.

\red{We should also mention some previous DFT studies on the interfacial and wetting behaviour of binary mixtures. This work is the background and has multiple connections to what we do here.
For one-component fluids, much insight can be gained by considering the Sullivan DFT model \cite{sullivan1979van}. This can be extended to consider binary mixtures and significant progress was made based on this approach \cite{sullivan1982interfacial, da1983adsorption, hadjiagapiou1985adsorption, ding1989some}.
Another worthwhile DFT-based approach is to make the sharp-kink approximation (i.e.\ assuming step-like interfacial density profiles), which facilitates deriving many useful results for the interfacial thermodynamics that then lead to the easy mapping out of wetting phase diagrams \cite{dietrich1986order, dietrich1989classification, chen2006bulk, kim2021wetting}.}

This paper is structured as follows: First, in Sec.~\ref{sec:bindingPotential}, we give an overview of the thermodynamics of liquids adsorbed on solid substrates and, in particular, the additional considerations required for the case that the liquid is a binary mixture. Then, in Sec.~\ref{sec:3}, we introduce the lattice DFT we use to model binary liquid mixtures. In Sec.~\ref{sec:4} we briefly explain how the bulk fluid phase behaviour depends on the state of the system (temperature, pressure and chemical potentials) and also on the strength of the pair interaction potentials between the different particles in the mixture. We display examples of phase diagrams, showing binodal curves for various systems under consideration. In Sec.~\ref{sec:5} we start our discussion of properties of the inhomogeneous liquids, considering first the density profiles of the two different species at the vapour-liquid interface, which must be obtained in order to calculate the liquid-vapour surface tension. After this, in Sec.~\ref{sec:6}, we move on to consider the behaviour of miscible binary liquids at solid substrates, displaying density profiles and the corresponding binding potentials $g(\Gamma_A,\Gamma_B)$, showing how these depend on the strength of the attraction between the substrate and the two different liquids. In Sec.~\ref{sec:7} we present density profiles and binding potentials for the case of immiscible liquids at a wall. In these cases, the binding potential can be a multi-valued function, with different branches corresponding to the different possible configurations of the system. In Sec.~\ref{sec:8} we present results corresponding to the case where the two adsorptions are equal $\Gamma_A=\Gamma_B$, to illustrate further the multi-valued nature of $g(\Gamma_A,\Gamma_B)$. In Sec.~\ref{sec:9} we show results from calculations where we allow the liquid density profiles to vary both in a direction parallel to the wall, as well as perpendicular to it. We present examples of various different possible droplet configurations, although the range of possible behaviours is rather large and so our results in this section are not intended to be comprehensive. Finally, in Sec.~\ref{sec:conc} we close with a few concluding remarks.

\section{Surface thermodynamics and binding potentials}\label{sec:bindingPotential}

\begin{figure}[t]
\centering

\includegraphics[width=0.95\columnwidth]{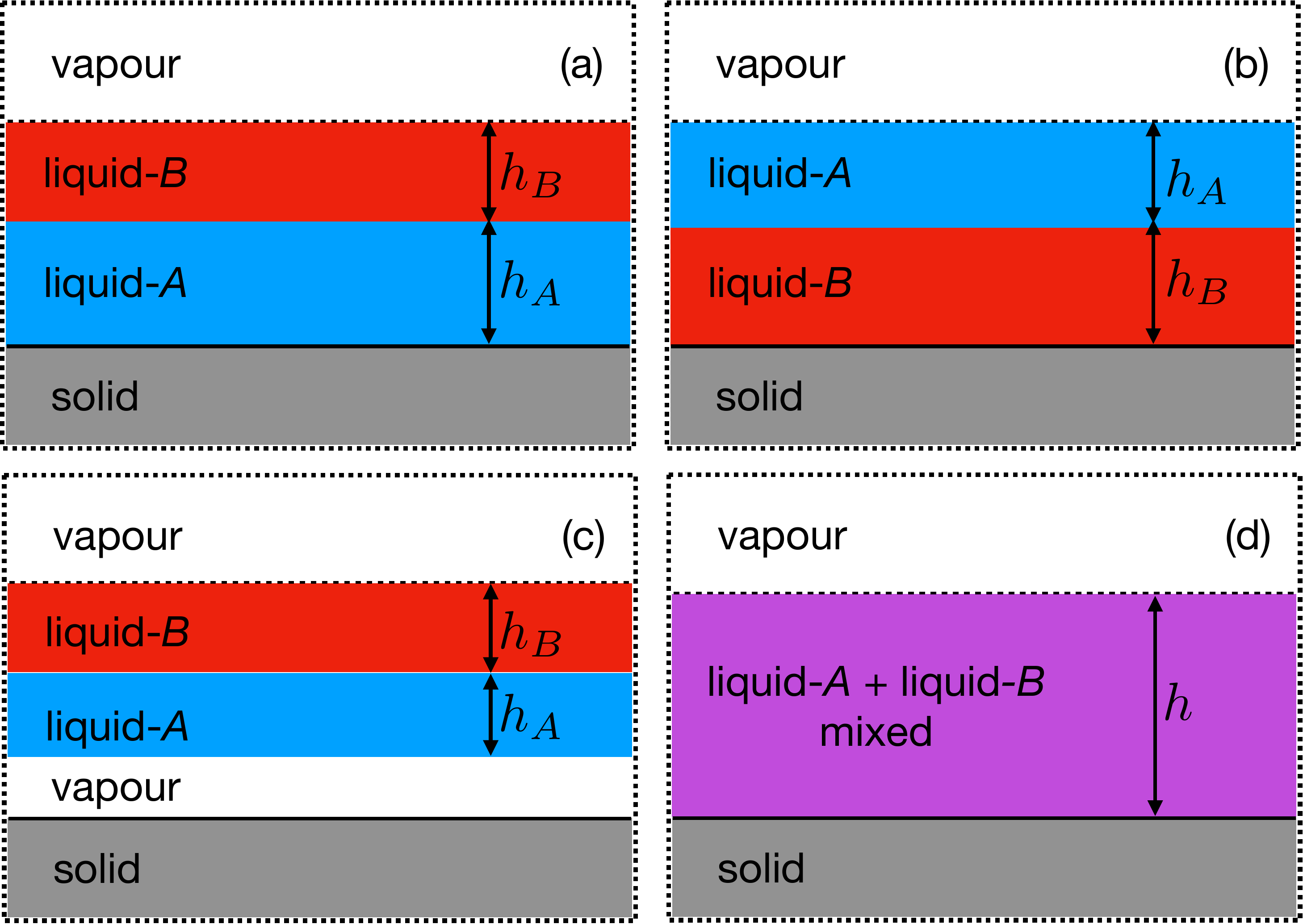}

\caption{Illustration of possible configurations of the system when two different liquids are deposited on a solid substrate, assuming that the density distributions vary only in the vertical $z$-direction, perpendicular to the wall. Panel (a) shows the case of two immiscible liquids, with liquid-$A$ in contact with the wall, while (b) shows the case where liquid-$B$ is in contact with the wall. Panel (c) shows the case where the wall is solvophobic and so a film of the vapour intrudes between the liquid-$A$ and the wall. A configuration where $A$ and $B$ are interchanged is also possible. Panel (d) shows the case where the two liquids are miscible and so cover the wall together as a mixture.}
\label{hthickness}
\end{figure}


Although our interest lies in two-component mixtures, we first discuss the case of a one-component liquid, allowing us to establish the fundamental ideas surrounding binding potentials.
We begin by considering the excess grand potential for a system with a film of liquid of thickness $h$ on a planar solid surface \cite{schick1990introduction, dietrich88}
\begin{equation}
\omega_{ex}(h) \equiv \frac{\Delta \Omega}{{\cal A}} = \gamma_{sl} + \gamma_{lv} + g(h)+h\Delta p_{lv},
\label{eq:excess grand energy_one_comp}
\end{equation}
where $\Delta\Omega=\Omega-\Omega_0$, with $\Omega$ the grand potential of the system, $\Omega_0=-p_v V$ the grand potential of the vapour phase in a container of the same volume $V$ but with no surface ($p_v$ is the pressure of the vapour), and ${\cal A}$ the area of the substrate's surface. The terms on the right-hand side are the solid-liquid interfacial tension $\gamma_{sl}$, the liquid-vapour interfacial tension $\gamma_{lv}$, $g$ is the binding potential, and the pressure difference $\Delta p_{lv}=p_l-p_v$, where $p_l$ is the pressure in the liquid film on the surface. Note, here we treat $h$ and $\Gamma$ as being essentially interchangeable.

When the chemical potential of the system $\mu$ is the value at vapour-liquid phase coexistence $\mu_{coex}$, then $\Delta p_{lv}=0$. 
Using the Gibbs-Duhem relation, when the system is close to coexistence, we find that $h\Delta p_{lv}=\Gamma \Delta \mu$, where $\Delta \mu =(\mu-\mu_{coex})$ \cite{dietrich88, chacko2017hydrophobicity}, which gives an alternative form for the last term in Eq.~\eqref{eq:excess grand energy_one_comp} that is more useful in the context of the present DFT calculations. 

Note that Eq.~\eqref{eq:excess grand energy_one_comp} can be considered as defining the binding potential: $g(h)$ is whatever is left when all the other terms on the right hand side are subtracted from the excess grand potential per unit area. Defining it this way makes clear that it has the property $g(h)\to0$ in the limit $h\to\infty$.

We now consider the same situation, but generalised to the case where the system contains a two-component mixture of particles of species $A$ and $B$. Examples of typical interfacial wetting behaviours are shown in Fig.~\ref{hthickness}. The top panels correspond to the cases when the two species demix (including in the bulk), exhibiting a coexistence between a phase rich in species-$A$ particles, which we refer to as `liquid-$A$' and a phase rich in species-$B$ particles, which we refer to as `liquid-$B$'. The top left panel (a) corresponds to the case where the film of liquid-$A$ is at the wall with the liquid-$B$ layer above it, whereas the top right panel (b) corresponds to the case where the $B$-rich layer is at the wall with the $A$-rich layer above it. The bottom left panel (c) also corresponds to the demixing situation, but now there is a layer of vapour at the wall with layers of liquids $A$ and $B$ above it. This configuration is generally metastable, but is sometimes relevant, e.g.\ when considering bubbles under the liquid film. We say more about this situation below. The bottom right panel (d) corresponds to the mixing situation.

For the demixing case in Fig.~\ref{hthickness}(a), when liquid-$A$ is at the wall, we can generalise Eq.~\eqref{eq:excess grand energy_one_comp} and write the excess grand potential of the system as \red{\cite{luengo2021lifshitz}}:
 \begin{eqnarray}
\omega_{A} &\equiv& \frac{\Omega_A - \Omega_0}{{\cal A}}\nonumber \\
&=& \Gamma_A\Delta\mu_A+\Gamma_B\Delta \mu_B + \gamma_{sA} + \gamma_{AB}+ \gamma_{Bv}\nonumber\\
&&+g_{A}(\Gamma_{A}, \Gamma_{B}),
\label{eq:excess grand energy_two_liquid-A}
\end{eqnarray}
where the subscript on the variables $\omega_A$, $g_A$ and $\Omega_A$ is there to remind us which phase is closest to the wall. The three interfacial tensions are the wall-liquid-$A$ tension $\gamma_{sA}$, the liquid-$A$-liquid-$B$ interfacial tension $\gamma_{AB}$ and the liquid-$B$-vapour tension $\gamma_{Bv}$. We also define the chemical potential differences $\Delta\mu_A=\mu_A-\mu_A^{coex}$ and $\Delta\mu_B=\mu_B-\mu_B^{coex}$, where the values of the two chemical potentials at bulk phase coexistence are $\mu_A^{coex}$ and $\mu_B^{coex}$ for the species-$A$ and species-$B$ particles, respectively. As for the one-component case, we can view this equation as defining the binding potential $g_A(\Gamma_{A}, \Gamma_{B})$.
Analogously to the one-component case, when $\Gamma_{A}\to\infty$ and $\Gamma_{B}\to\infty$, then we have $g_A\to0$.

For the case illustrated in Fig.~\ref{hthickness}(b), when liquid-$B$ is the phase in contact with the wall, we can write an entirely analogous equation to the one above:
 \begin{eqnarray}
\omega_{B} &\equiv& \frac{\Omega_B - \Omega_0}{{\cal A}} \nonumber\\
&=& \Gamma_A\Delta\mu_A+\Gamma_B\Delta \mu_B + \gamma_{sB} + \gamma_{AB}+ \gamma_{Av}\nonumber\\
&&+g_{B}(\Gamma_{A}, \Gamma_{B}),
\label{eq:excess grand energy_two_liquid-B}
\end{eqnarray}
where $g_B$ is the binding potential in this case with liquid-$B$ at the wall and it also has the property that when $\Gamma_{A}\to\infty$ and $\Gamma_{B}\to\infty$, then $g_B\to0$.

An obvious question is how to determine which of the two states in Fig.~\ref{hthickness}(a) or (b) is preferred by the system.  We consider the special case of when the chemical potentials both equal their values at the triple point, where all three phases (liquid-$A$, liquid-$B$ and the vapour) are in bulk phase-coexistence, then $\Delta\mu_A=\Delta\mu_B=0$. Also assuming that all of the films are thick enough that both $g_A\approx0$ and $g_B\approx0$, results in simplified forms for Eqs.~\eqref{eq:excess grand energy_two_liquid-A} and \eqref{eq:excess grand energy_two_liquid-B}: 
\begin{equation}
\omega_{A}^{\rm coex}(\Gamma_A\to\infty,\Gamma_B\to\infty) = \gamma_{sA} + \gamma_{AB}+ \gamma_{Bv},
\end{equation}
and
\begin{equation}
\omega_{B}^{\rm coex}(\Gamma_A\to\infty,\Gamma_B\to\infty) = \gamma_{sB} + \gamma_{AB}+ \gamma_{Av}.
\end{equation}
Whichever of these two quantities is the lowest corresponds to the state preferred by the system -- i.e.\ this determines whether the substrate prefers to have liquid-$A$ or liquid-$B$ in contact with it. This ordering is, of course, determined by how the wall interacts with the two different species. We define the constant $\delta$ as the difference between these two values, i.e.\
\begin{equation}
\delta = \left( \gamma_{sA} - \gamma_{sB} \right)+ \left(\gamma_{Bv} - \gamma_{Av}\right).
\end{equation}
Below in Sec.~\ref{sec:7} we calculate the grand potential for a range of different substrates. In order to compare results for given substrate properties corresponding to the various different configurations in Fig.~\ref{hthickness}, we plot the interface potential
\begin{equation}
\label{omega_delta}
\tilde{\omega}^\alpha \equiv \omega_{\alpha} - (\gamma_{sA} + \gamma_{AB} + \gamma_{Bv}),
\end{equation}
for all the possible configurations of the liquid, and where $\alpha=A,B$ denotes the phase at the wall. Thus, when it is liquid-$A$ at the wall, then
\begin{equation}
\label{omega_a}
\tilde{\omega}^A = g_{A}
\end{equation}
and when it is liquid-$B$ at the wall, then
\begin{equation}
\label{omega_b}
\tilde{\omega}^B = g_{B} - \delta.
\end{equation}
We can plot these two quantities in the same figure and the offset between the two cases, $\tilde{\omega}^A(\Gamma_A,\Gamma_B)$ and $\tilde{\omega}^B(\Gamma_A,\Gamma_B)$, is $\delta$ when $\Gamma_A$ and $\Gamma_B$ are large. The magnitude of $\delta$ is an indicator for how much the wall prefers to have liquid-$A$ or liquid-$B$ next to it. 

We now discuss the remaining two cases in Fig.~\ref{hthickness}. For systems where a (metastable) film of the vapour can exist between the wall and the liquid, as illustrated in Fig.~\ref{hthickness}(c), we can write the excess grand potential as
 \begin{eqnarray}
\omega^{v} &\equiv& \frac{\Omega_v - \Omega_0}{{\cal A}}  \nonumber \\
&=& \Gamma_A\Delta\mu_A+\Gamma_B\Delta \mu_B+\gamma_{sv} + \gamma_{vA} + \gamma_{AB} + \gamma_{Bv}\nonumber \\
&&+ g_{v}(\Gamma_{A}, \Gamma_{B}).
\label{eq:excess grand energy_two_liquid-v}
\end{eqnarray}
Similarly to the above case, we define
\begin{equation}
\label{omega_v}
\tilde{\omega}^v = g_v - \delta_{v},
\end{equation}
where $\delta_{v}$ is given by
\begin{equation}
\delta_{v} = \left(\gamma_{sA} - \gamma_{sv} \right) - \gamma_{vA}.
\label{eq:delta_v}
\end{equation}
The value of this quantity tells us how favourable it is for such a film of the vapour to intrude at the wall.  We note that the analogous case to that in 
Fig.~\ref{hthickness}(c) where the $A$ and $B$ liquid phases are interchanged is also possible, and similar statements to those above hold.

\red{Finally, we consider the case when the mixing of the two species of particles is favourable, which is the case illustrated in Fig.~\ref{hthickness}(d), and where Eq.~\eqref{eq:excess grand energy_one_comp} applies, since there is just one liquid phase together with the vapour phase present in the system.
Here, there is no longer the need to consider two separate adsorptions (or two separate film thicknesses) and the system reverts to an effective one component system and the standard formulation for such systems applies, though differences should emerge on approaching a tricritical point \cite{indekeu2022wetting}. Nonetheless, one can still vary the two adsorptions $\Gamma_A$ and $\Gamma_B$ independently, which is entirely equivalent to varying the bulk concentrations of the two species in the single liquid phase in contact with the wall.
In this situation, we plot below in Sec.~\ref{sec:6} the binding potential surface $g(\Gamma_A,\Gamma_B)$.} Following this, in Sec.~\ref{sec:7} we then move on to consider the more complex case of liquids that demix at substrates.
However, before we present these results, having in this section briefly reviewed the relevant thermodynamics of binary liquid mixtures at interfaces, we next describe the specific microscopic lattice-DFT that we use to calculate the quantities introduced above.

\section{Lattice DFT for binary mixtures}\label{sec:3}

We model binary liquid mixtures by discretising these systems of interacting particles onto a \red{three-dimensional (3D)} cubic lattice with lattice spacing $\sigma$. We set $\sigma = 1$, defining our unit of length. Each site on the lattice is labelled by index $\mathbf{i}$, where $\mathbf{i} = (i,j,k)$ is the 3D discrete position vector. We define $l_{\mathbf{i}}^{A}$ and $l_{\mathbf{i}}^{B}$ as the occupation numbers for particles of species-$A$ and species-$B$ at site $\mathbf{i}$, respectively. Thus, if site $\mathbf{i}$ is occupied by a particle of species-$A$, then $l_{\mathbf{i}}^{A} = 1$ and if the site is unoccupied, then $l_{\mathbf{i}}^{A} = 0$. Similarly, $l_{\mathbf{i}}^{B} = 1$ or $0$ depending on whether or not the site $\mathbf{i}$ is occupied by a particle of species-$B$. We also assume that a lattice site cannot be occupied by both types at the same time, i.e.\ $l_{\mathbf{i}}^{A} + l_{\mathbf{i}}^{B} = 0$ or $1$, but not $2$.  We model the total energy (Hamiltonian) of the system $E$ in any given configuration $\{l_{\mathbf{i}}^{A},l_{\mathbf{i}}^{B}\}$ by the following sum \cite{woywod2003phase, chalmers2017modelling, robbins2011modelling, archer2010dynamical}:
\begin{align}
\label{eq:Hamiltonian_liquids}
E &= - \sum_{\mathbf{i},\mathbf{j}}\left(\frac{1}{2} \varepsilon_{\mathbf{ij}}^{AA} l_{\mathbf{i}}^A l_{\mathbf{j}}^A + \varepsilon_{\mathbf{ij}}^{AB} l_{\mathbf{i}}^A l_{\mathbf{j}}^B + \frac{1}{2} 
\varepsilon_{\mathbf{ij}}^{BB} l_{\mathbf{i}}^B l_{\mathbf{j}}^B\right) 
\nonumber
\\
&- \mu_{A} \sum_{\mathbf{i}} l_{\mathbf{i}}^{A} - \mu_{B} \sum_{\mathbf{i}} l_{\mathbf{i}}^{B} + \sum_{\mathbf{i}} \Phi_{\mathbf{i}}^{A} l_{\mathbf{i}}^A 
 + \sum_{\mathbf{i}} \Phi_{\mathbf{i}}^{B} l_{\mathbf{i}}^B.
\end{align}
The first term, a sum over pairs of lattice sites, is the contribution from particle-particle interactions.
The overall interaction between pairs of species-$A$ particles at sites $\mathbf{i}$ and $\mathbf{j}$ is determined by the discretised pair potential $\varepsilon_{\mathbf{ij}}^{AA} = \varepsilon_{AA} c_{\mathbf{ij}}$, which depends on the distance between lattice sites, $|\mathbf{i}-\mathbf{j}|$. The parameter $\varepsilon_{AA}$ governs the overall strength. Similarly, $\varepsilon_{\mathbf{ij}}^{AB} = \varepsilon_{AB} c_{\mathbf{ij}}$ is the interaction tensor between species-$A$ and species-$B$ particles, with strength determined by the parameter $\varepsilon_{AB}$, and $\varepsilon_{\mathbf{ij}}^{BB} = \varepsilon_{BB} c_{\mathbf{ij}}$ is the interaction between pairs of species-$B$ particles, with $\varepsilon_{BB}$ determining the overall strength. Here $c_{\mathbf{ij}}$ is a dimensionless coefficient which decreases in value as the distance between the pairs of particles increases. There are various possible choices one could \red{make. For a 3D model, the following values are a good choice:}
\begin{equation}
\label{eq:c_ij_liquids}
\red{c_{\mathbf{ij}}  = 
  \begin{cases} 
   1 & \text{if }\mathbf{j}\in {NN \mathbf{i}}, \\
    \frac{3}{10}    & \text{if }\mathbf{j}\in {NNN \mathbf{i}}, \\
    \frac{1}{20}    & \text{if }\mathbf{j}\in {NNNN \mathbf{i}}, \\
        0   & \text{otherwise},
         \end{cases}}
\end{equation}
where $NN\mathbf{i}$, $NNN\mathbf{i}$ and $NNNN\mathbf{i}$ denote the nearest neighbours of $\mathbf{i}$, next nearest neighbours of $\mathbf{i}$ and next-next nearest neighbours of $\mathbf{i}$, respectively. The choice of values in Eq.~(\ref{eq:c_ij_liquids}) is important: with these particular values any equilibrium droplets that form on the surface tend to have a hemispherical \red{shape \cite{kumar2004isotropic, chalmers2017modelling, chalmers2017dynamical}.} If, for example, one were to assume only nearest neighbour interactions, i.e.\ with $c_{\mathbf{ij}}=0$, for $\mathbf{j} \in {NNN \mathbf{i}}$ and ${NNNN \mathbf{i}}$, then rectangular shaped droplets are liable to be formed, particularly at low temperatures. The choice of $c_{\mathbf{ij}}$ in Eq.~(\ref{eq:c_ij_liquids}) minimises the dependence of the vapour-liquid surface tension on the orientation of the interface with respect to the lattice orientation. \red{In two dimensions (2D), the equivalent of Eq.~\eqref{eq:c_ij_liquids} is \cite{kumar2004isotropic}
\begin{equation}
\label{eq:c_ij_2D}
c_{\mathbf{ij}}  = 
  \begin{cases} 
   1 & \text{if }\mathbf{j}\in {NN \mathbf{i}}, \\
    \frac{1}{2}    & \text{if }\mathbf{j}\in {NNN \mathbf{i}},\\
        0   & \text{otherwise},
         \end{cases}
\end{equation}
used e.g.\ in the model discussed in Refs.~\cite{robbins2011modelling, archer2010dynamical}.}

The second and third terms in Eq.~(\ref{eq:Hamiltonian_liquids}) are the contributions from treating the system as being coupled to a reservoir, which is the vapour above the surface. The chemical potentials, $\mu_{A}$ and $\mu_{B}$, of species-$A$ and species-$B$ determine the rate at which liquid-$A$ and liquid-$B$ evaporate from (or condense onto) the surface, respectively.

The last two terms in Eq.~(\ref{eq:Hamiltonian_liquids}) are sums over all the lattice sites and give the contributions to the potential energy from the interactions with the surface, which exerts the external potentials $\Phi_{\mathbf{i}}^{A}$ and $\Phi_{\mathbf{i}}^{B}$ on species-$A$ and species-$B$, respectively. We model these as follows:
\begin{equation}
\label{eq:ext_pot_A}
\Phi_{\mathbf{i}}^{\alpha} = 
  \begin{cases} 
    \infty  & k < 1\\
    -\varepsilon_{w\alpha} & k = 1\\
    0   &  \text{otherwise},
        \end{cases}
\end{equation}
where $\alpha = A, B$ and $k$ is the perpendicular distance from the surface. Here $\varepsilon_{w\alpha}$ is the parameter which determines the interaction strength between the particles of species-$\alpha$ and the surface (or wall). The boundary condition for the particles at the wall is straight-forward: we set $l_{\mathbf{i}}^\alpha= 0$ for all lattice sites with $k < 1$ in Eq.~(\ref{eq:ext_pot_A}), where $k = 0$ is the position of the planar surface of the substrate, i.e.\ of the wall. For our DFT calculations, we use periodic boundary conditions in the $i$ and $j$ directions that are parallel to the wall. For the top boundary, we either assume that the system is closed, i.e.\ so that this is also a hard purely repulsive wall, or that the system is open so that it is an absorbing boundary, which is modelled by fixing the density on that boundary to that of the vapour with the corresponding chemical potentials. We set the top boundary sufficiently far from the wall, so that it  does not affect the resulting densities in the region of the wall.

To develop a statistical mechanical theory for the system we employ DFT, resulting in a theory for the average densities
\begin{align}
\rho_{\mathbf{i}}^A = \langle l_{\mathbf{i}}^A\rangle \,\,\,\,\text{and} \,\,\,\,\rho_{\mathbf{i}}^B = \langle l_{\mathbf{i}}^B\rangle,
\end{align}
which are the ensemble average densities at site $\mathbf{i}$,  i.e.\ $\langle\cdots\rangle$ denotes a statistical average. Making a mean-field approximation, the Helmholtz free energy for the binary lattice-gas is \cite{hughes2014introduction, chalmers2017dynamical}:
\begin{align}
\label{eq:free_energy_Binary}
F\left(\{\rho_{\mathbf{i}}^A\}, \{\rho_{\mathbf{i}}^B\}\right) 
&= k_{B}T \sum_{\mathbf{i}} \big[\rho_{\mathbf{i}}^A \ln{\rho_{\mathbf{i}}^A}
+ \rho_{\mathbf{i}}^B \ln{\rho_{\mathbf{i}}^B}
\nonumber \\
& + \left(1 - \rho_{\mathbf{i}}^A - \rho_{\mathbf{i}}^B \right) \ln{\left(1 - \rho_{\mathbf{i}}^A - \rho_{\mathbf{i}}^B\right)}
\big]
\nonumber
\\ 
& - \frac{1}{2}\sum_{\mathbf{i},\mathbf{j}} \varepsilon_{\mathbf{ij}}^{AA} \rho_{\mathbf{i}}^A \rho_{\mathbf{j}}^A - \sum_{\mathbf{i},\mathbf{j}}\varepsilon_{\mathbf{ij}}^{AB} \rho_{\mathbf{i}}^A \rho_{\mathbf{j}}^B \nonumber \\ 
& - \frac{1}{2} 
\sum_{\mathbf{i},\mathbf{j}}\varepsilon_{\mathbf{ij}}^{BB} \rho_{\mathbf{i}}^B \rho_{\mathbf{j}}^B 
\nonumber
\\ 
& + \sum_{\mathbf{i}} \left( \Phi_{\mathbf{i}}^{A} \rho_{\mathbf{i}}^A + \Phi_{\mathbf{i}}^{B} \rho_{\mathbf{i}}^B\right),
\end{align}
where $k_B$ is Boltzmann's constant and $T$ is the temperature. The above is a discretised DFT free energy for a binary mixture. A more accurate expression for $F$ can be obtained by following Refs.~\cite{maeritz2021density, maeritz2021droplet}, but for our present purposes the above mean-field approximation is sufficient. \red{When the densities are slowly varying, it is straightforward to map the above lattice-DFT onto a continuum-DFT with a gradient expansion approximation form \cite{robbins2011modelling}. The entropic terms in the second line of Eq.~\eqref{eq:free_energy_Binary} solely contribute a local term in the continuum limit and can  loosely be thought of as a `hard-sphere repulsion' term.} In the following sections, we minimize $F$ subject to constraints on the adsorptions of the two species, in the presence of a wall (i.e., with an external potential).  However, in the next section we first consider what happens in bulk mixtures, where no wall is present.

\section{Bulk mixture phase behaviour}\label{sec:4}

Without a wall, i.e.\ with external potentials $\Phi_{\mathbf{i}}^A = \Phi_{\mathbf{i}}^B = 0$, equilibrium states of the system are a uniform fluid with constant number densities $\rho_{\mathbf{i}}^A = \rho_{A}$ and $\rho_{\mathbf{i}}^B= \rho_{B}$. 
In this case, the sum over neighbours in the interaction terms in the Helmholtz free energy (\ref{eq:free_energy_Binary}) can be evaluated explicitly. The integrated interaction tensor is \red{$\sum_{\mathbf{j}}c_{\mathbf{ij}} = {\cal S}$ for all $\mathbf{i}$, where in 3D, ${\cal S}=10$, and in 2D, ${\cal S}=6$, so we have $a_{AA} = {\cal S} \varepsilon_{AA}$, $a_{AB} = {\cal S} \varepsilon_{AB}$, $a_{BB} = {\cal S} \varepsilon_{BB}$} as the integrated strengths of the pair interaction potentials. From Eq.~\eqref{eq:free_energy_Binary} the Helmholtz free energy per unit volume, $f = F/V$, where $V$ is the volume of the system, is given by:
\begin{align}
\label{eq:free_energy_Binary_per_lattice}
f & = k_{B}T\big[\rho_A \ln{\rho_A} + \rho_B \ln{\rho_B} \nonumber \\
& \qquad \qquad + \left(1 - \rho_A - \rho_B \right) \ln{\left(1 - \rho_A - \rho_B\right)} \big]
\nonumber
\\ 
& - \frac{1}{2} a_{AA}\rho_{A}^2 - a_{AB} \rho_{A} \rho_{B} - \frac{1}{2} a_{BB} \rho_{B}^2.
\end{align}
From this we may calculate the spinodal, the locus where $(\partial^2 f / \partial \rho_A^2)(\partial^2 f / \partial \rho_B^2)-(\partial^2 f / \partial \rho_A\partial\rho_B)^2 = 0$. 
This spinodal defines the boundary of the region of the phase diagram where the system is unstable, and density fluctuations in a uniform system spontaneously grow, leading to phase separation. For temperatures where two-phase coexistence equilibria can occur, the binodal curve gives the coexisting density values.
States in the phase diagram outside the binodal are stable, and no phase separation occurs. The binodal is calculated by equating the chemical potentials, temperatures and pressures in each of the coexisting phases.  For this we can use Eq.~(\ref{eq:free_energy_Binary_per_lattice}) since thermodynamic quantities such as the chemical potentials, $\mu_A$ and $\mu_B$, and pressure, $P$, may be obtained using the following relations:\red{  
\begin{eqnarray}
\mu_A = \frac{\partial f}{\partial \rho_A},\quad\mu_B = \frac{\partial f}{\partial \rho_B},\quad
P = -\left.\frac{\partial (fV)}{\partial V}\right|_{N_A,N_B,T}.
\end{eqnarray}
}
These give:
\begin{align}
\mu_A & = k_BT\left( \ln \rho_A - \ln\left(1 - \rho_A - \rho_B\right)\right) - a_{AA}\rho_A - a_{AB}\rho_B,
\\
\mu_B & = k_BT\left( \ln \rho_B - \ln\left(1 - \rho_A - \rho_B\right)\right) - a_{AB}\rho_A - a_{BB}\rho_B,
\\
 P & = - k_BT \ln\left(1 - \rho_A - \rho_B\right) \nonumber \\
 & - \frac{1}{2} a_{AA}\rho_A^2 - a_{AB}\rho_A\rho_B - \frac{1}{2} a_{BB}\rho_B^2,
\end{align}
where we have used the fact that in a uniform system the densities are $\rho_A = N_A/V$ and $\rho_B = N_B/V$ where $N_A$ and $N_B$ are the total number of  particles of each species in the system.

\begin{figure}[t!]
\begin{center}
\includegraphics[width=\columnwidth]{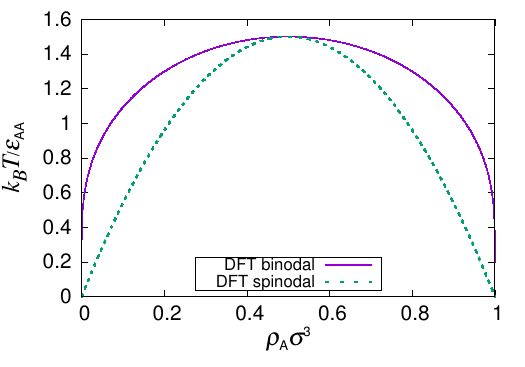}
\caption{Binodal curve for the coexisting densities of the liquid and vapour phases for the pure one-component lattice fluid, calculated using DFT. We also display the spinodal. }
\label{fig:phase_digram_liquids_a}
\end{center}
\end{figure}

For the one component pure $A$-type with no particles of species-$B$ (i.e.\ $\rho_B = 0$), we can use the symmetry of the Hamiltonian (\ref{eq:Hamiltonian_liquids}) to simplify the calculation of the binodal curve for the coexisting vapour and liquid densities of species-$A$ particles. This allows us to observe that if $\rho_A$ is the density of the liquid at coexistence then $(1 - \rho_A)$ is the density of the coexisting vapour. On equating the pressure in the two phases, \red{for ${\cal S}=6$ (2D case)} we obtain the following equation for the binodal:
\begin{equation}
\frac{k_BT}{\varepsilon_{AA}} = \frac{3(2\rho_A - 1)}{\ln\left[\rho_A/(1-\rho_A)\right]}.
\end{equation}
This has a maximum at $\rho_A = 0.5$ which corresponds to a critical temperature of $k_BT = 1.5\varepsilon_{AA}$. Figure~\ref{fig:phase_digram_liquids_a} shows a plot of this binodal curve together with the spinodal.

\begin{figure}
\begin{center}
\includegraphics[width=\columnwidth]{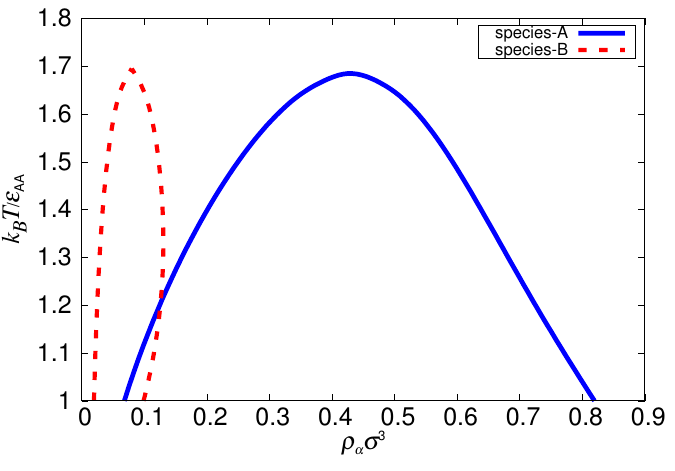}
\caption{Binodal curves for the binary mixture with $\mu_B/ \varepsilon_{AA}=  -4.25$, $\varepsilon_{AB}/ \varepsilon_{AA}=  0.825$, $\varepsilon_{BB}/ \varepsilon_{AA}=  0.65$. Since $\mu_B$ is low, there is only a relatively small amount of species-$B$ in each of the coexisting liquid and vapour phases.}
\label{fig:phase_digram_liquids_c}
\end{center}
\end{figure}

\begin{figure}[t!]
\begin{center}
\includegraphics[width=\columnwidth]{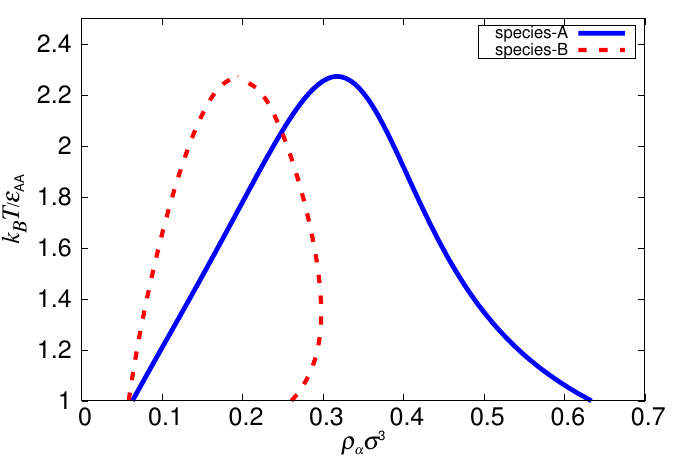}
\caption{Binodal curves for the binary mixture with $\mu_B/ \varepsilon_{AA}=  -3.25$, $\varepsilon_{AB}/ \varepsilon_{AA}=  0.825$, $\varepsilon_{BB}/ \varepsilon_{AA}=  0.65$. In this case, the two liquids are somewhat miscible, and so the densities of both in each of the two coexisting phases are sizeable.}
\label{fig:phase_digram_liquids_b}
\end{center}
\end{figure}

\begin{figure}
\begin{center}
\includegraphics[width=\columnwidth]{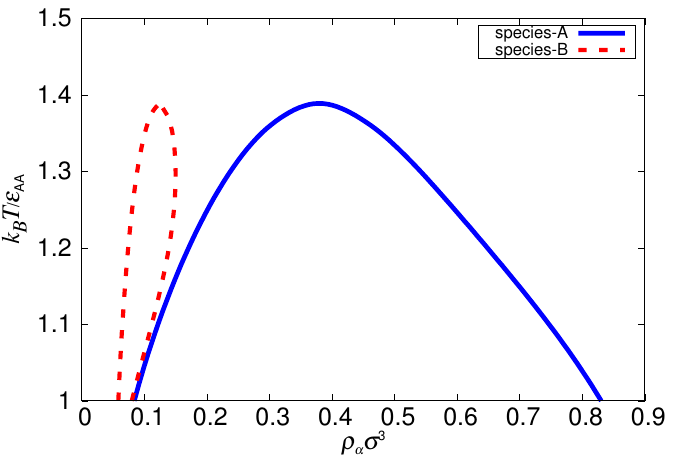}
\caption{Binodal curves for the binary mixture with $\mu_B/ \varepsilon_{AA}=  -3.25$, $\varepsilon_{AB}/ \varepsilon_{AA}=  0.575$, $\varepsilon_{BB}/ \varepsilon_{AA}=  0.65$. The value of $\varepsilon_{AB}$ here is smaller than the case in Fig.~\ref{fig:phase_digram_liquids_b}, so this case corresponds to a pair of liquids that are less miscible, which is why the density of the species-$B$ particles is lower in this case.}
\label{fig:phase_digram_liquids_d}
\end{center}
\end{figure}

For the binary mixture to exhibit two fluid phases that coexist in thermodynamic equilibrium, four conditions must be satisfied. We denote these two phases as (i) the low-density phase $(LDP)$ and (ii) the high-density phase $(HDP)$, although when we consider some liquid-liquid demixing cases, the total density difference between the two coexisting phases can be very small or even zero. We must then solve the following set of simultaneous equations:
\begin{align}
T^{LDP} & = T^{HDP},
\label{eq: first_condition}
\\
\mu_{A}^{LDP} & = \mu_{A}^{HDP},
\label{eq: second_condition}
\\
\mu_{B}^{LDP} & = \mu_{B}^{HDP},
\label{eq: third_condition}
\\
P^{LDP} & = P^{HDP},
\label{eq: fourth_condition}
\end{align}
 for the densities of the two different species in the two different phases, $\rho_{A}^{LDP}$, $\rho_{B}^{LDP}$, $\rho_{A}^{HDP}$ and $\rho_{B}^{HDP}$ (four unknowns).
 
 The first equation above is trivial to enforce -- we simply pick the temperature of interest and require that it is the same in both phases. We are then left with three equations for four unknowns. To make progress, we then fix the chemical potential of species-$B$ to some specified value, which we denote as $\eta$, so that Eq.~\eqref{eq: third_condition} above decouples to give us two new equations: $\mu_{B}^{LDP} = \eta$ and $\mu_{B}^{HDP} = \eta$. We can then solve these two equations together with Eqs.~(\ref{eq: second_condition}) and (\ref{eq: fourth_condition}) for the four density values: $\rho_{A}^{LDP}$, $\rho_{B}^{LDP}$, $\rho_{A}^{HDP}$ and $\rho_{B}^{HDP}$. Thus, we get four equations for the four unknowns \cite{robbins2011modelling}. Solving these over a range of different temperatures and values of $\eta$ allows us to determine the phase diagram.

Figures \ref{fig:phase_digram_liquids_c} and \ref{fig:phase_digram_liquids_b} show the binodals for the mixture of species-$A$ and species-$B$ particles for the case when $\varepsilon_{AB}/ \varepsilon_{AA}=  0.825$, $\varepsilon_{BB}/ \varepsilon_{AA}=  0.65$ and for different values of the species-$B$ chemical potential $\mu_B$. We see that as $\mu_B$ is increased, the density of species-$B$ increases in both phases and in fact become the majority species for large enough $\mu_B$. Note that Fig.\ \ref{fig:phase_digram_liquids_a} can be considered to be the $\mu_B =  -\infty$ case in this sequence with varying $\mu_B$, where, of course, $\rho_B= 0$ in both coexisting phases.

In Fig.\ \ref{fig:phase_digram_liquids_d} we show results for a case where $\varepsilon_{AB}$ is less than both $\varepsilon_{BB}$ and $\varepsilon_{AA}$, in contrast to the cases in Figs.~\ref{fig:phase_digram_liquids_c} and \ref{fig:phase_digram_liquids_b}, where $\varepsilon_{AB} = (\varepsilon_{BB} +\varepsilon_{AA})/2$. In this case, we find that species-$B$ does not mix well with species-$A$ and when $\mu_B/ \varepsilon_{AA} =  -3.25$ (a low value), the density of the species-$B$ particles in both coexisting phases is low.

\section{Density profiles at the free interface}\label{sec:5}

\begin{figure}[htbp]
\centering
  \includegraphics[width=0.9\columnwidth]{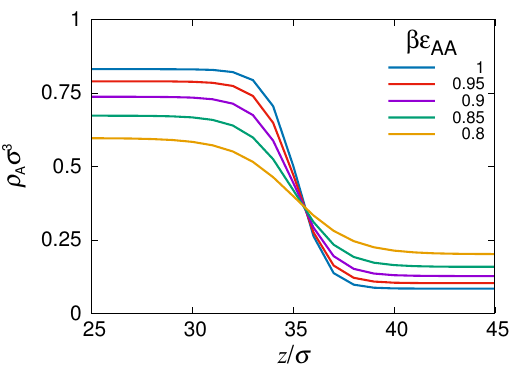}
 \includegraphics[width=0.9\columnwidth]{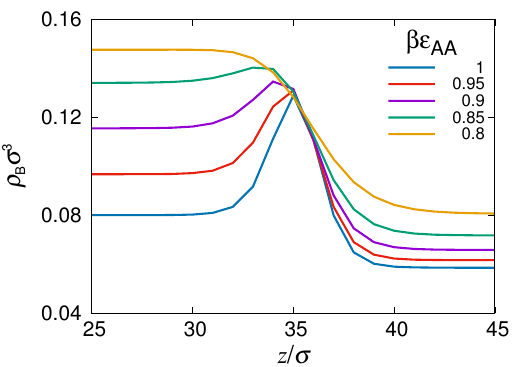}
 \caption{Liquid-$A$ density profiles (top) and liquid-$B$ density profiles (bottom) at the free liquid-vapour interface for $\varepsilon_{AB}/\varepsilon_{AA} = 0.575$,  $\varepsilon_{BB}/\varepsilon_{AA} = 0.65$ at different values of  $\beta\varepsilon_{AA}$, as indicated in the key. The corresponding bulk fluid phase diagram is shown in Fig.\ \ref{fig:phase_digram_liquids_d}.}
\label{fig:interface_binary}
\end{figure}

Having determined the bulk fluid phase behaviour, we now briefly consider the interface between the coexisting phases without a wall.
At the planar interface between the vapour and the liquid phases, the density profiles vary only in the direction perpendicular to the interface. We assume that the index varying in the direction perpendicular to the interface is $k$, recall that $\mathbf{i} = (i, j, k)$, so that the lattice sites are along the $z$-axis at $z=j\sigma$. The density profiles are calculated by minimising the grand potential \cite{evans1979nature, hansen2013theory}
\begin{equation}
\Omega = F - \mu_A \sum_{\mathbf{i}} \rho_{\mathbf{i}}^A - \mu_B \sum_{\mathbf{i}} \rho_{\mathbf{i}}^B,
\label{eq: minimising the grand potential_binary}
\end{equation}
where the Helmholtz free energy $F$ is given by Eq.~\eqref{eq:free_energy_Binary} and the chemical potentials $\mu_A$ and $\mu_B$ are set to be the values at which vapour-liquid phase coexistence occurs. In order to produce an interface, we set the density to the liquid (HDP) value
for small $z$, and to the vapour (LDP) value for large $z$.

As mentioned, here the density profiles only vary in one direction, which means that the full 3D DFT formalism can be averaged into 1D, significantly decreasing the computational complexity of the problem.  However, for consistency with later results, which in some cases also vary in two directions, we compute the densities here in 2D \red{(and set ${\cal S}=6$)}, with the constraint that there is no variation in the direction parallel to the interface.

In Fig.~\ref{fig:interface_binary} we display the density profiles of the two species with $\mu_B/\varepsilon_{AA} = -3.25$, $\varepsilon_{AB}/\varepsilon_{AA} = 0.575$, $\varepsilon_{BB}/\varepsilon_{AA} = 0.65$ for various temperatures, which corresponds to the bulk phase diagram in Fig.~\ref{fig:phase_digram_liquids_d}. We observe that when the temperature increases (i.e.\ the value $\beta\varepsilon_{AA}$ decreases) the total density difference between the two coexisting phases decreases. Note that for larger values of $\beta\varepsilon_{AA}$ (lower temperatures) there is a peak in the species-$B$ density profile at the interface, indicating that this species is enriched at the interface, i.e.\ it has a slight surfactant-like behaviour. This occurs at lower temperatures when mixed at low concentrations into the species-$A$ rich liquid. To calculate the various interfacial tensions, e.g.\ in Eq.~\eqref{eq:excess grand energy_two_liquid-B}, we can substitute the corresponding interfacial density profiles (such as the profiles in Fig.~\ref{fig:interface_binary}) into Eq.~\eqref{eq: minimising the grand potential_binary} to calculate the grand potential of the system with the interface $\Omega$, and then subtracting the corresponding bulk value $\Omega_0=-pV$ (with no interface) allows us to calculate the interfacial tension as $\gamma_{Av}=(\Omega-\Omega_0)/{\cal A}$, where ${\cal A}$ is the area of the interface. A similar calculation including the wall allows to determine all the other interfacial tensions \cite{hughes2014introduction}. The wall-liquid tensions of course depend on where we define the Gibbs dividing surface \cite{RowlinsonWidom1982}, but as long as it does not change between calculations, everything remains consistent.

\section{DFT results for miscible liquids at a wall}\label{sec:6}

\begin{figure}[t!]
\begin{center}
\includegraphics[width=0.9\columnwidth]{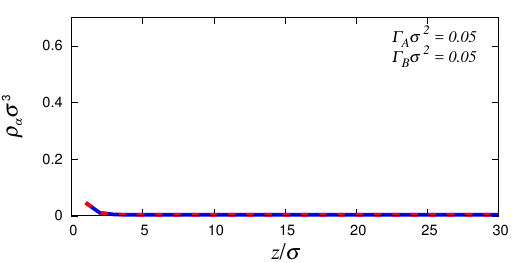}
\includegraphics[width=0.9\columnwidth]{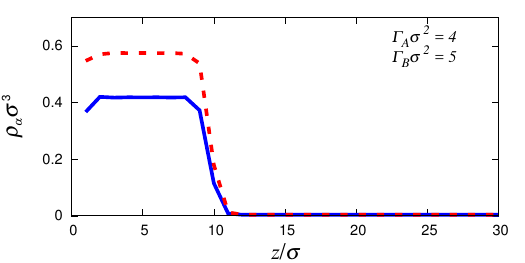}
\includegraphics[width=0.9\columnwidth]{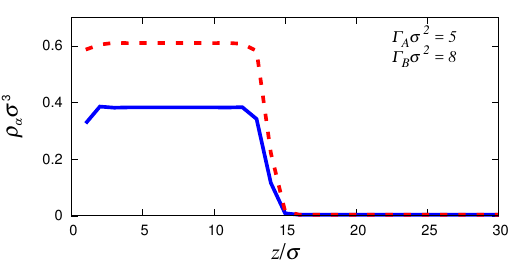}
\includegraphics[width=0.9\columnwidth]{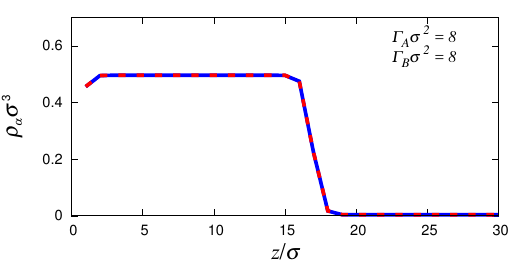}
\caption{Density profiles for a symmetric binary miscible liquid mixture at a planar wall, for varying values of the adsorptions $\Gamma_A$ and $\Gamma_B$ for species-$A$, and species-$B$, respectively, as indicated in the key of each panel. The solid (blue) line is for species-$A$ and the dashed (red) line is for species-$B$. The particle interaction parameters are $\varepsilon_{AA}=\varepsilon_{BB}$ and $\varepsilon_{AB}/\varepsilon_{AA} = 1.2$. The temperature of the system is $k_B{T}/\varepsilon_{AA} = 0.67$ and the chemical potentials are $\mu_A/\varepsilon_{AA}=\mu_B/\varepsilon_{AA}=-5.6$, so that the densities of each species in the bulk gas phase are $\rho_A\sigma^3=\rho_B\sigma^3\approx0.004$. The wall attraction strength parameters are $\varepsilon_{wA}/\varepsilon_{AA} =\varepsilon_{wB}/\varepsilon_{AA} = 1$. The binding potential corresponding to these profiles is displayed in Fig.~\ref{fig:10}.}
\label{fig:7}
\end{center}
\end{figure}

\begin{figure}[t!]
\begin{center}
\includegraphics[width=0.9\columnwidth]{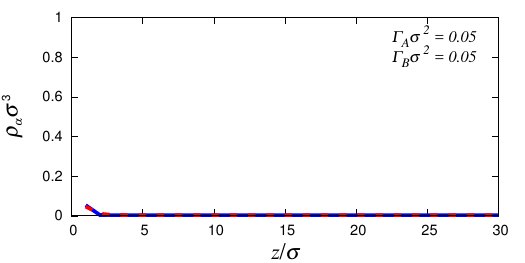}
\includegraphics[width=0.9\columnwidth]{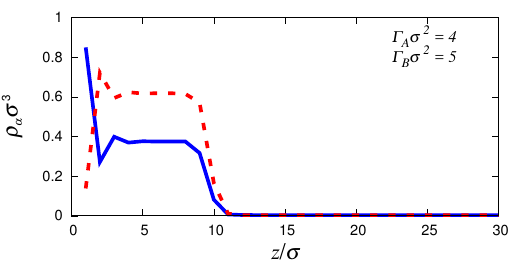}
\includegraphics[width=0.9\columnwidth]{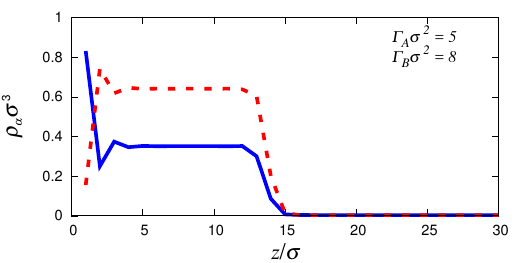}
\includegraphics[width=0.9\columnwidth]{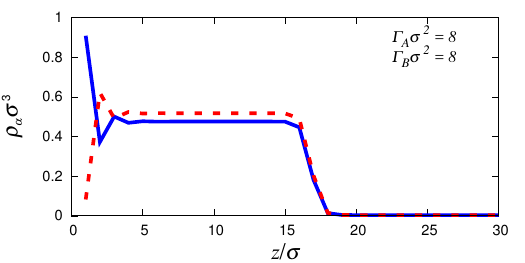}
\caption{Density profiles for a system exactly the same as that in Fig.~\ref{fig:7}, except here we increase the attraction between the species-$A$ particles and the wall to $\varepsilon_{wA}/\varepsilon_{AA} = 3.9$ (from the previous value of 1). The binding potential corresponding to these profiles is displayed in Fig.~\ref{fig:11}.}
\label{fig:8}
\end{center}
\end{figure}

\begin{figure}[t!]
\begin{center}
\includegraphics[width=0.9\columnwidth]{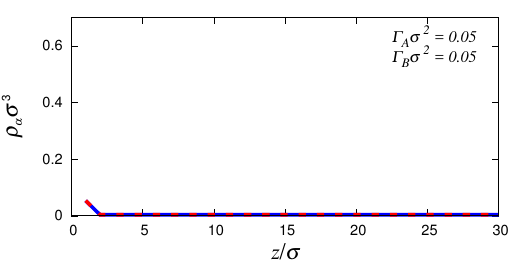}
\includegraphics[width=0.9\columnwidth]{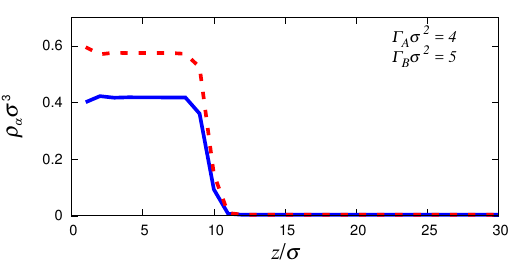}
\includegraphics[width=0.9\columnwidth]{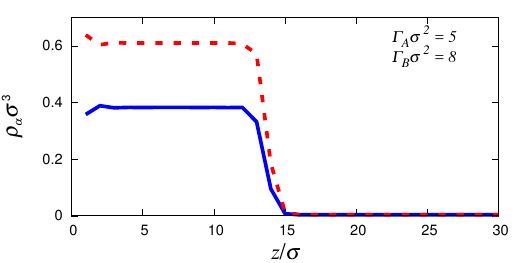}
\includegraphics[width=0.9\columnwidth]{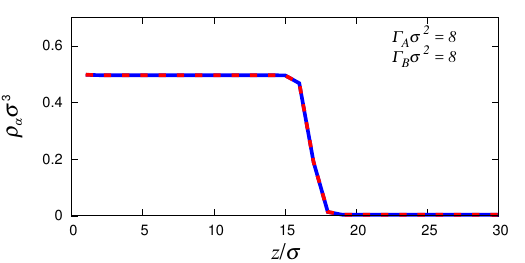}
\caption{Density profiles for a system exactly the same as that in Fig.~\ref{fig:7}, except here we increase the attraction between the wall and both species of particles to $\varepsilon_{wA}/\varepsilon_{AA} = \varepsilon_{wB}/\varepsilon_{AA} = 3.9$ (from the previous value of 1). Note that the main differences between the results here and those in Fig.~\ref{fig:7} are in the vicinity of the wall. The binding potential corresponding to these profiles is displayed in Fig.~\ref{fig:12}.}
\label{fig:9}
\end{center}
\end{figure}

In this section we present results for the density profiles and binding potential $g(\Gamma_A,\Gamma_B)$ for miscible liquids at solid substrates. This is done by minimising the free energy \eqref{eq:free_energy_Binary} subject to the constraint that the two adsorptions $\Gamma_A$ and $\Gamma_B$ are the specified values and then repeating over the desired range of values of $(\Gamma_A,\Gamma_B)$. As mentioned in the introduction, we use the constrained minimisation algorithm developed in Ref.~\cite{hughes2015liquid}, generalised to binary mixtures. 

For the binary mixtures considered here, the two main factors which determine the final configuration of the system are (i) the strengths of the external potentials attracting the particles towards the planar wall and (ii) the strength of the interactions between the particles of the two different species -- in particular, whether the cross interaction attraction strength $\varepsilon_{AB}$ is large enough to prevent demixing or not.

We consider first the case of a symmetric miscible binary mixture with $\varepsilon_{BB}=\varepsilon_{AA}$ and $\varepsilon_{AB}/\varepsilon_{AA}=1.2$. We set the temperature $k_{B}{T}/\varepsilon_{AA} = 0.67$. Since $\varepsilon_{AB}>\varepsilon_{AA} = \varepsilon_{BB}$ (i.e.\ the particles are more strongly attracted to the opposite species than they are to their own kind), the mixture does not exhibit liquid-liquid demixing and only exhibits vapour-liquid phase separation. We now consider various different combinations of the wall attraction strength parameters. Note that for a planar wall, the equilibrium density profiles vary only with the distance from the wall. 

For the results shown in Fig.~\ref{fig:7}, we set $\varepsilon_{wA} = \varepsilon_{wB} = \varepsilon_{AA}$ and display some typical density profiles.  These range from the top panel with $\Gamma_{A}\sigma^2 = \Gamma_{B}\sigma^2 = 0.05$ corresponding to a small excess of each species adsorbed at the wall (essentially no liquid at the wall), to the bottom panel where $\Gamma_{A}\sigma^2 = \Gamma_{B}\sigma^2 = 8$, corresponding to a large excess adsorbed at the wall (a thick film of liquid). 

In Figs.~\ref{fig:8} and \ref{fig:9} we show corresponding density profiles for different values of $\varepsilon_{wA}$ and $\varepsilon_{wB}$; comparing these two figures with Fig.~\ref{fig:7} allows to see the influence of the relative wall attraction strengths on the density profiles. In Fig.~\ref{fig:8} we increase the attraction between only species-$A$ and the wall, whereas in Fig.~\ref{fig:9} we increase both species-wall attractions.

In Fig.~\ref{fig:8}, we display the fluid density profiles for the case where we increase $\varepsilon_{wA}$ to the value $\varepsilon_{wA}/\varepsilon_{AA} = 3.9$. The bulk phase behaviour is the same as the previous case (Fig.~\ref{fig:7}), so the two species of particles prefer to remain mixed in the liquid phase. However, particles of species-$A$ now have a much greater tendency than species-$B$ particles to be at the wall, because $\varepsilon_{wA} > \varepsilon_{wB}$. This is reflected in the density profiles of the species-$A$ particles, which have a sharp peak near to contact with the wall. The species-$A$ particles at the wall therefore exclude the species-$B$ particles from the wall surface and so there is a corresponding drop in the species-$B$ density profiles near to the wall. 

The density profiles for the final case we consider in this section are displayed in Fig.~\ref{fig:9}. The fluid is exactly the same miscible mixture as considered in Fig.~\ref{fig:7}, but now we increase the value of both $\varepsilon_{wA}/\varepsilon_{AA}$, and $\varepsilon_{wB}/\varepsilon_{AA}$ to $3.9$. The system is now symmetric again with respect to the wall interactions. The stronger attraction results in the liquid strongly wetting the wall. 


\begin{figure}[t!]

(a)
\includegraphics[width=0.9\columnwidth]{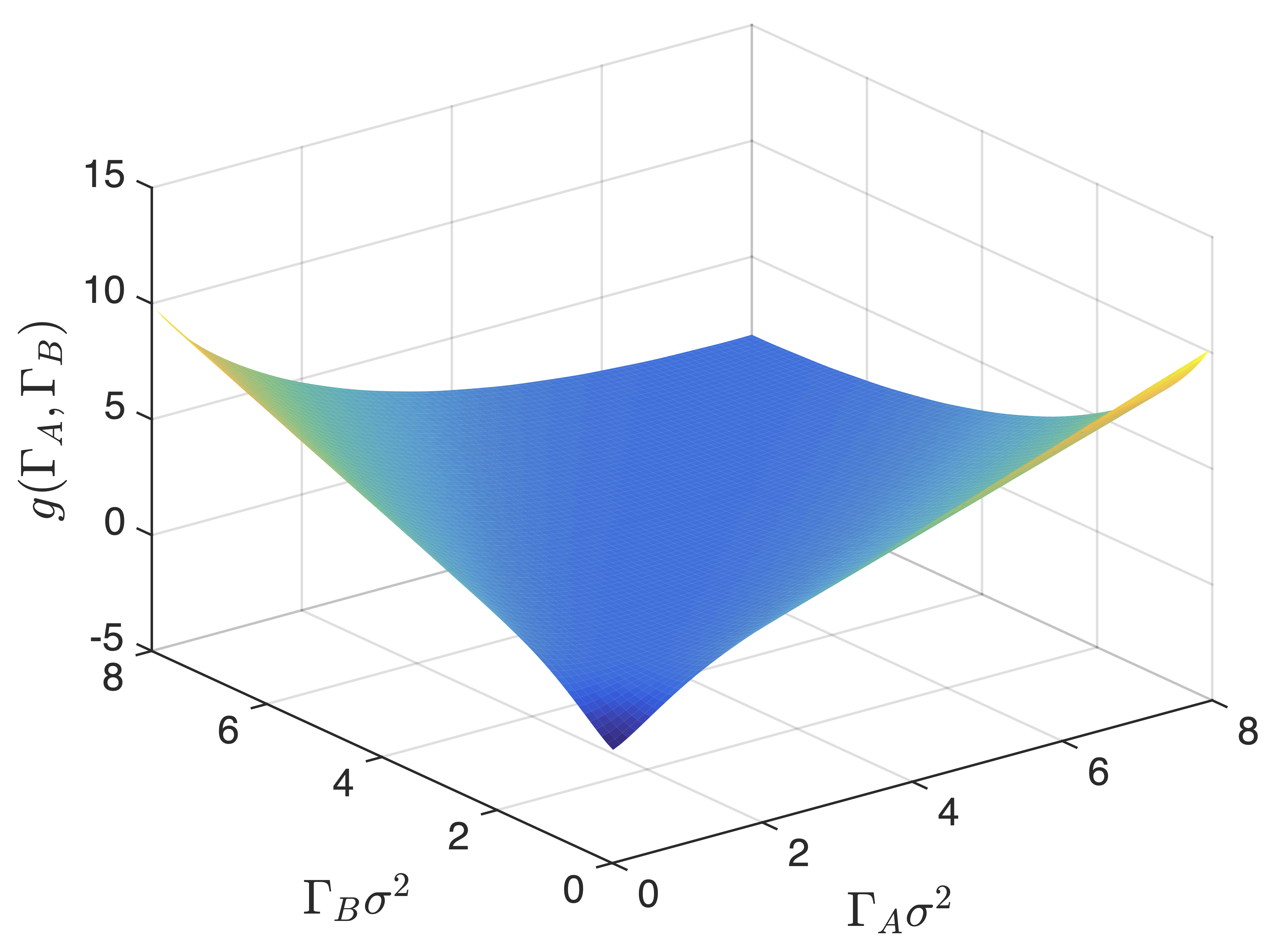}

(b)
\includegraphics[width=0.9\columnwidth]{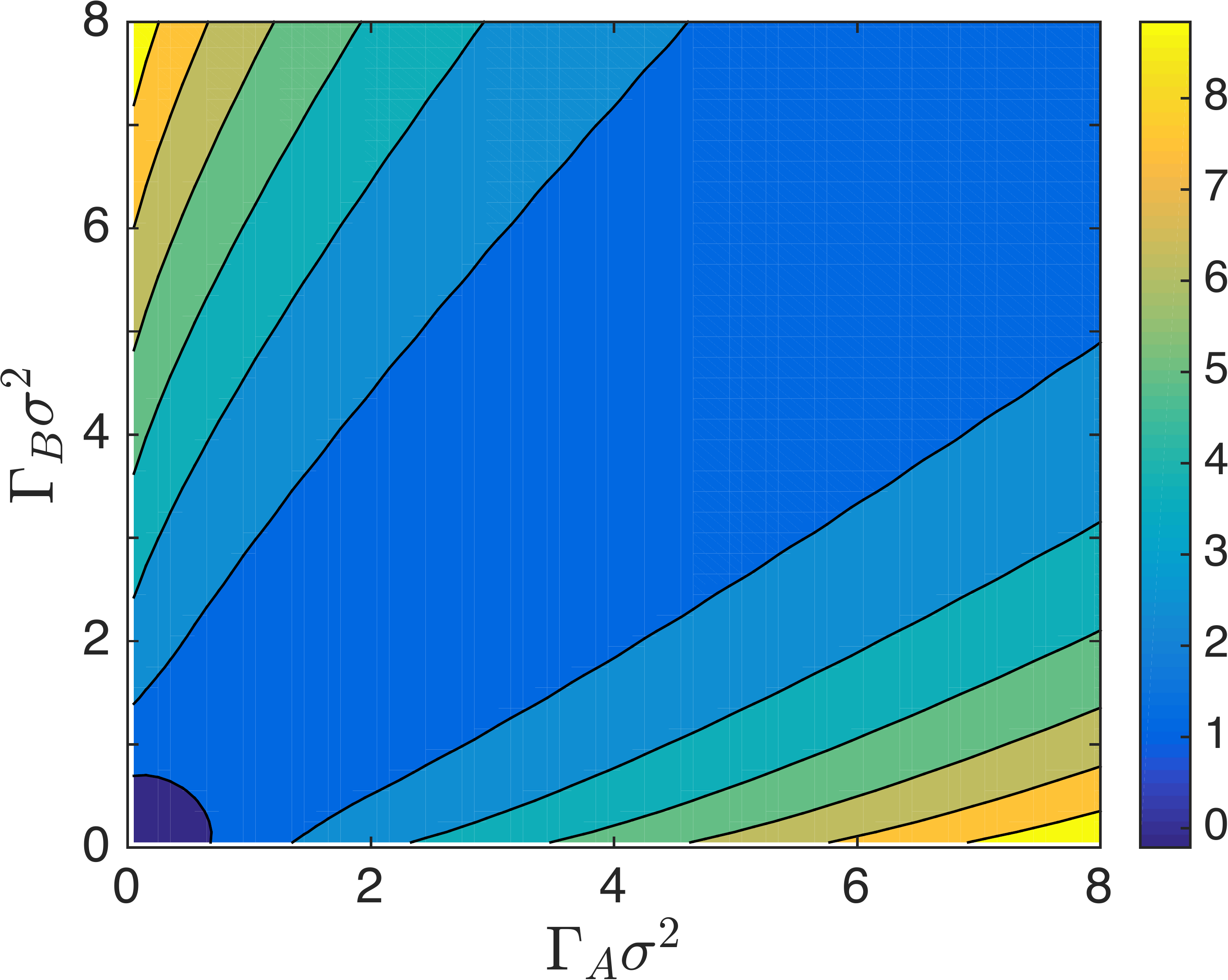}

\caption{The binding potential $g(\Gamma_{A}, \Gamma_{B})$ corresponding to the system in Fig.~\ref{fig:7} displayed as a surface plot in panel (a), and as a contour plot in panel (b). This is for a symmetric binary mixture that does not wet the wall, as can be seen from the fact that the global minimum is near the origin.}
\label{fig:10}
\end{figure}

\begin{figure}[t!]
(a)
\includegraphics[width=0.9\columnwidth]{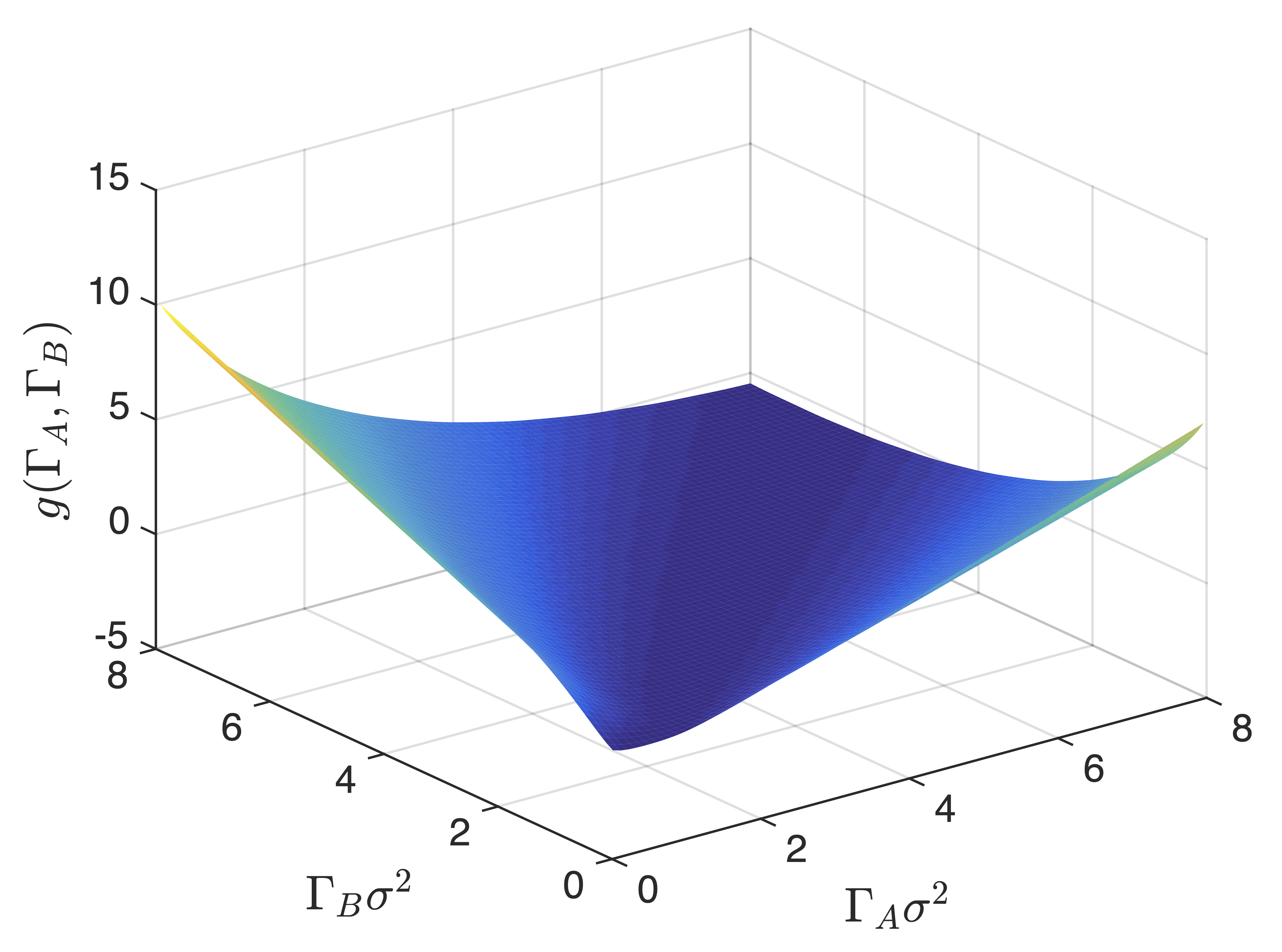}

(b)
\includegraphics[width=0.9\columnwidth]{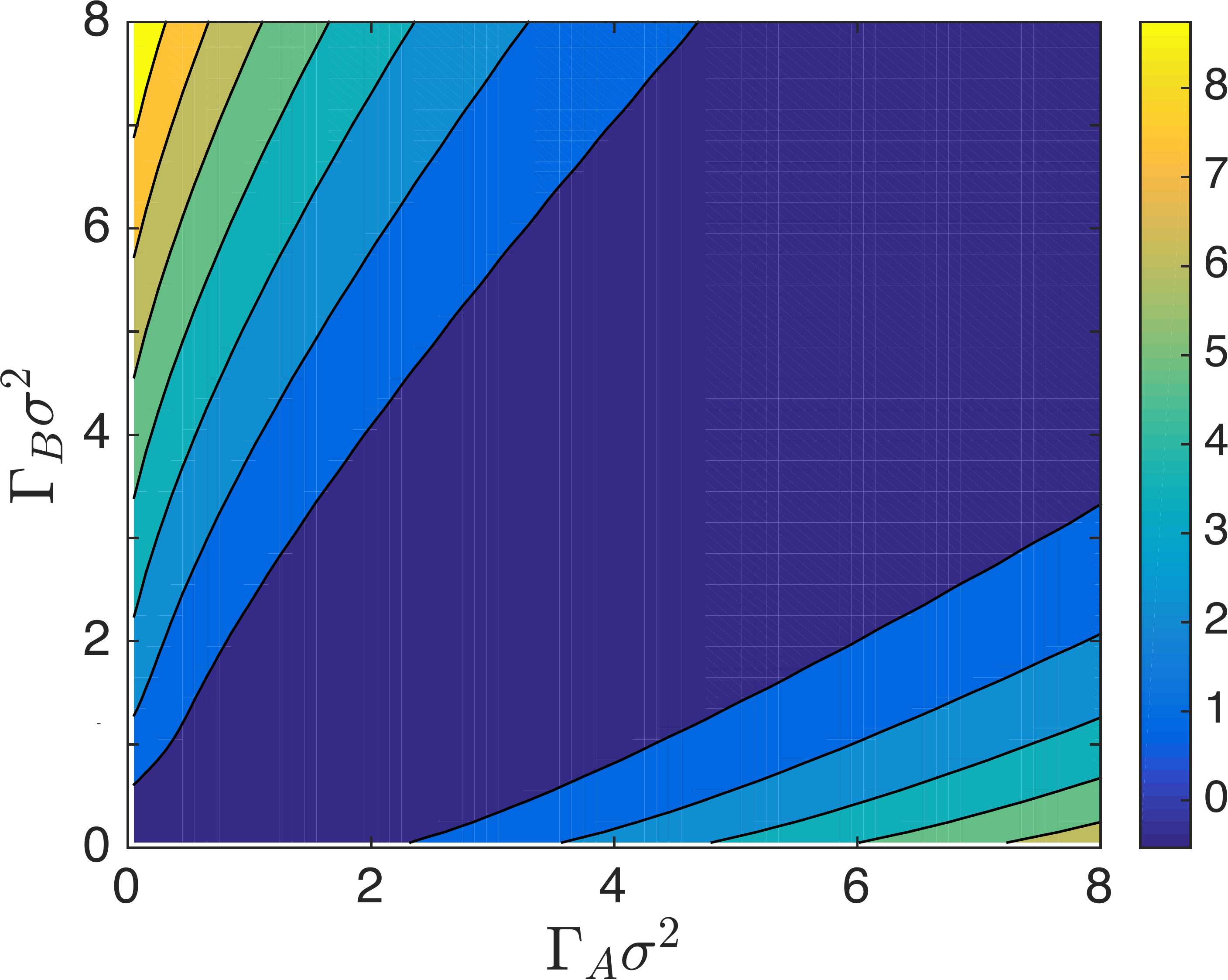}

\caption{The binding potential $g(\Gamma_{A}, \Gamma_{B})$ corresponding to the system in Fig.~\ref{fig:8} displayed as a surface plot in panel (a), and as a contour plot in panel (b). The stronger attraction between the wall and species-$A$ (compared to Figs.~\ref{fig:7} and \ref{fig:10}) leads to the fluid wetting the wall, as indicated by the fact that the global minimum of the binding potential is at $(\Gamma_A,\Gamma_B)\to(\infty,\infty)$.}
\label{fig:11}
\end{figure}

\begin{figure}[t!]

(a)
\includegraphics[width=0.9\columnwidth]{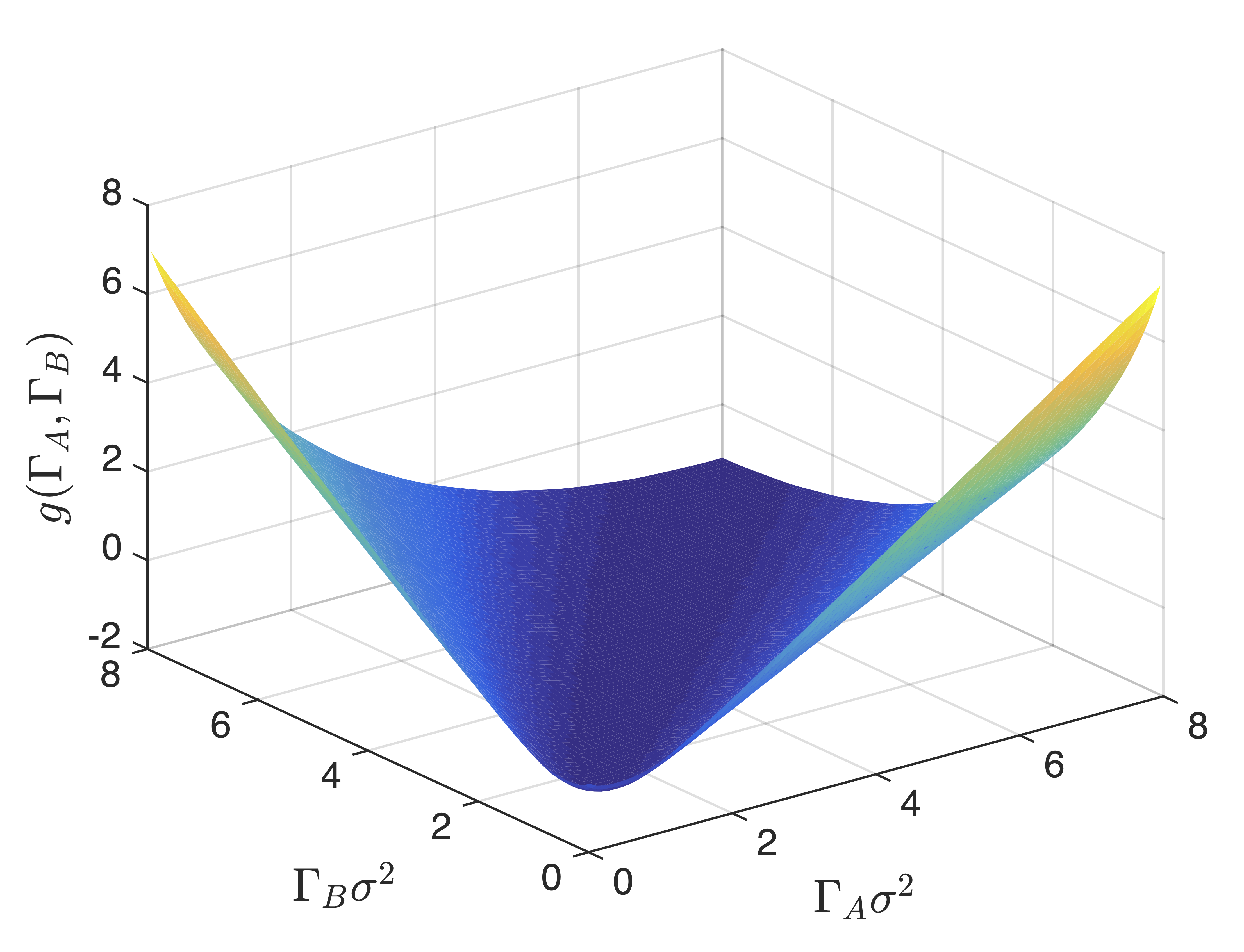}

(b)
\includegraphics[width=0.9\columnwidth]{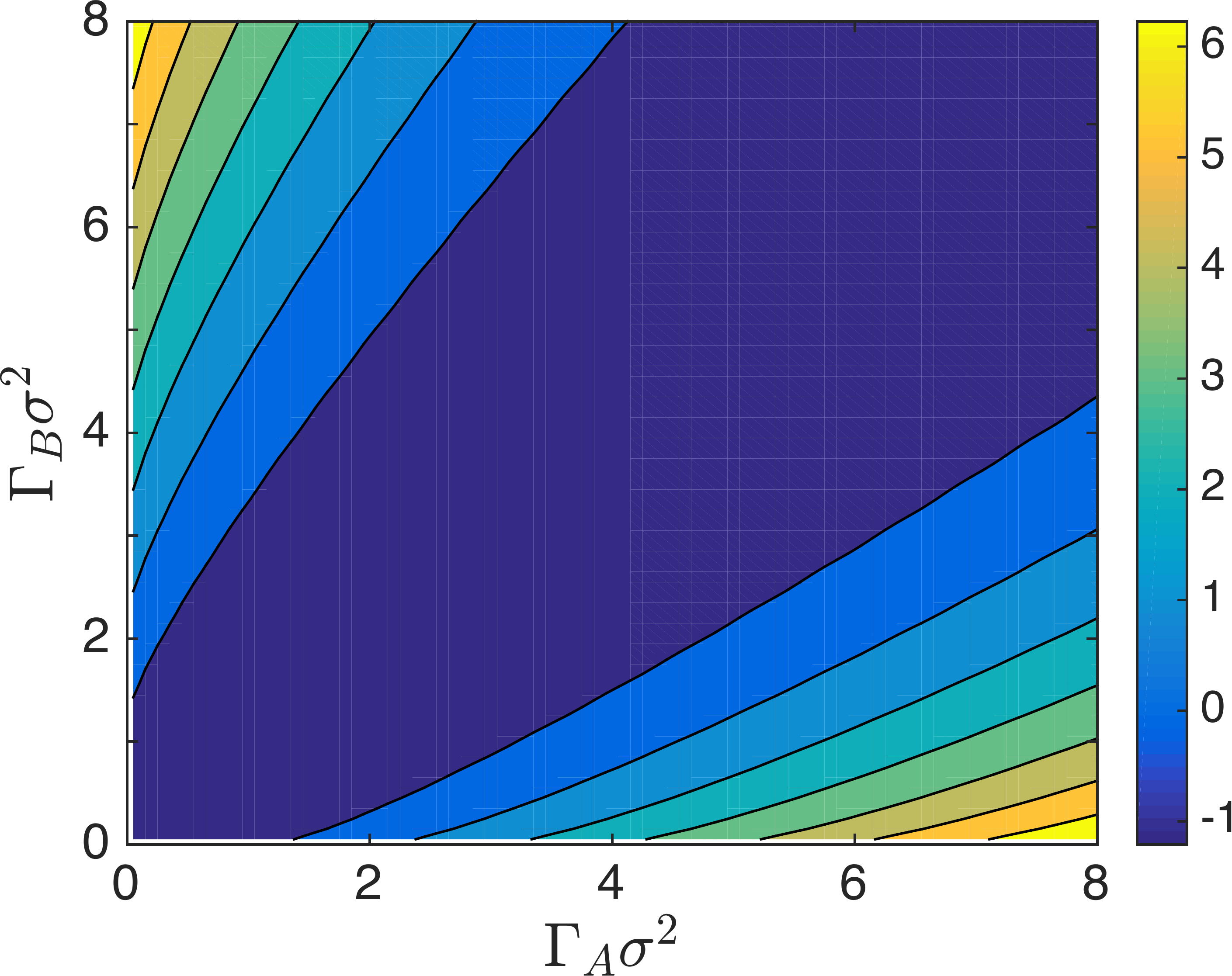}

\caption{The binding potential $g(\Gamma_{A}, \Gamma_{B})$ corresponding to the system in Fig.~\ref{fig:9} displayed as a surface plot in panel (a), and as a contour plot in panel (b). The stronger attraction between the wall and both species (compared to Figs.~\ref{fig:7} and \ref{fig:10}) leads to the fluid wetting this wall, as indicated by the fact that the global minimum of the binding potential is at $(\Gamma_A,\Gamma_B)\to(\infty,\infty)$.}
\label{fig:12}
\end{figure}

We now move on to discuss the binding potential $g(\Gamma_{A},\Gamma_{B})$ corresponding to the density profiles in Figs.~\ref{fig:7}, \ref{fig:8}, and \ref{fig:9}.  
These are shown in Figs.~\ref{fig:10}, \ref{fig:11}, and \ref{fig:12} where panel (a) shows the binding potential as a surface in $(\Gamma_{A}, \Gamma_{B},g)$ space, and panel (b) shows the corresponding contour plot in the $(\Gamma_{A}, \Gamma_{B})$ plane. Both plots use the colour bar from panel (b).

The particular shape of the binding potential surface in Fig.~\ref{fig:10} comes as a consequence of species-$A$ and species-$B$ having the same affinity for the wall (leading to symmetry across the $\Gamma_{A}=\Gamma_{B}$ line) and since, in this case, the liquid does not wet the wall; it is partially wetting, with a finite contact angle $\theta>0$. This latter fact can be seen from the observation that the binding potential has its global minimum value near (but not exactly at) the origin, where $(\Gamma_A,\Gamma_B)\approx(0,0)$. Note that even for small negative values of $\Gamma_A$ or $\Gamma_B$ the value of $g$ rapidly becomes very large, somewhat similar to the case for the one-component liquid discussed in Ref.~\cite{hughes2015liquid}. Also, for fixed total adsorption $\Gamma_A+\Gamma_B$ the binding potential is minimised where $\Gamma_{A} = \Gamma_{B}$, indicating that a mixing configuration is preferred.

The corresponding binding potential for the system shown in Fig.~\ref{fig:8} is plotted in Fig.~\ref{fig:11}, where we see the influence of the different attraction of the two species to the wall. Note that now the liquid phase wets the wall, facilitated by the species-$A$ particles strongly adsorbing at the wall thereby forming an enrichment layer. This can be seen from the fact that the global minimum of the binding potential is at $(\Gamma_A,\Gamma_B)\to(\infty,\infty)$, rather than near the origin. Note also that due to the asymmetry of the wall potentials, the binding potential surface is no longer symmetric around the line $\Gamma_A=\Gamma_B$.

Finally, the binding potential corresponding to Fig.~\ref{fig:9} is plotted in Fig.~\ref{fig:12}, which demonstrates that both species have the same affinity for the wall. Note that the global minimum of the binding potential is once again at $(\Gamma_A,\Gamma_B)\to(\infty,\infty)$, indicating that also in this case the liquid wets the wall.

\section{Results for immiscible liquid mixtures at a wall}\label{sec:7}

\begin{figure}[t!]
\begin{center}
\includegraphics[width=0.9\columnwidth]{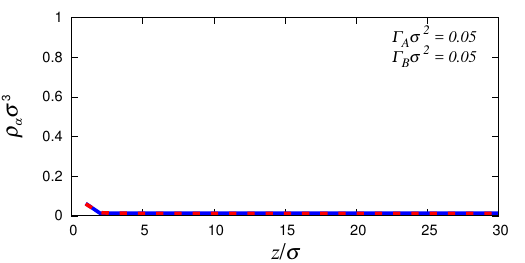}
\includegraphics[width=0.9\columnwidth]{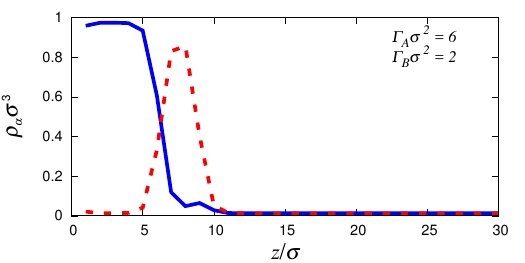}
\includegraphics[width=0.9\columnwidth]{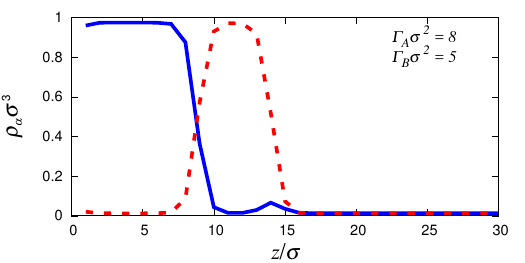}
\includegraphics[width=0.9\columnwidth]{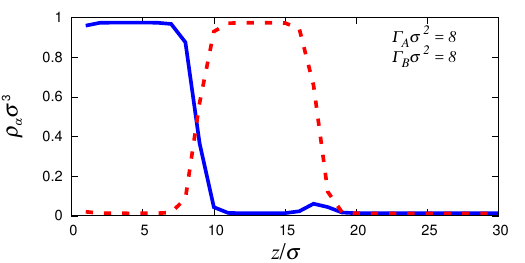}
\caption{Density profiles for a binary liquid mixture at a wall. The mixture exhibits bulk liquid-liquid demixing as well as gas-liquid phase separation. The density profiles at the wall reflect this, forming two distinct liquid layers, one above the other, in contact with the vapour. The profiles are for varying values of the adsorptions $\Gamma_A$ and $\Gamma_B$, as indicated in the keys. The solid (blue) line is for species-$A$ and the dashed (red) line is for species-$B$. These results are for a symmetric mixture with $\varepsilon_{BB}=\varepsilon_{AA}$ and $\varepsilon_{AB}/\varepsilon_{AA}= 0.75$. The temperature is $k_{B}{T}/\varepsilon_{AA} = 0.67$ and the chemical potentials are $\mu_A/\varepsilon_{AA} = \mu_B/\varepsilon_{AA} = -4.5$, so that the bulk vapour densities are $\rho_A\sigma^3=\rho_B\sigma^2 = 0.0128$ and the densities of the two species in the liquid phases are 0.974 and 0.0128. The wall interaction strength parameters are $\varepsilon_{wA}/\varepsilon_{AA}= 2.5$ and $\varepsilon_{wB}/\varepsilon_{AA}= 1.5$, so in this case the wall favours the liquid-$A$ phase.}
\label{fig:de_mixing_case_down}
\end{center}
\end{figure}

\begin{figure}[t!]
(a)
\includegraphics[width=0.9\columnwidth]{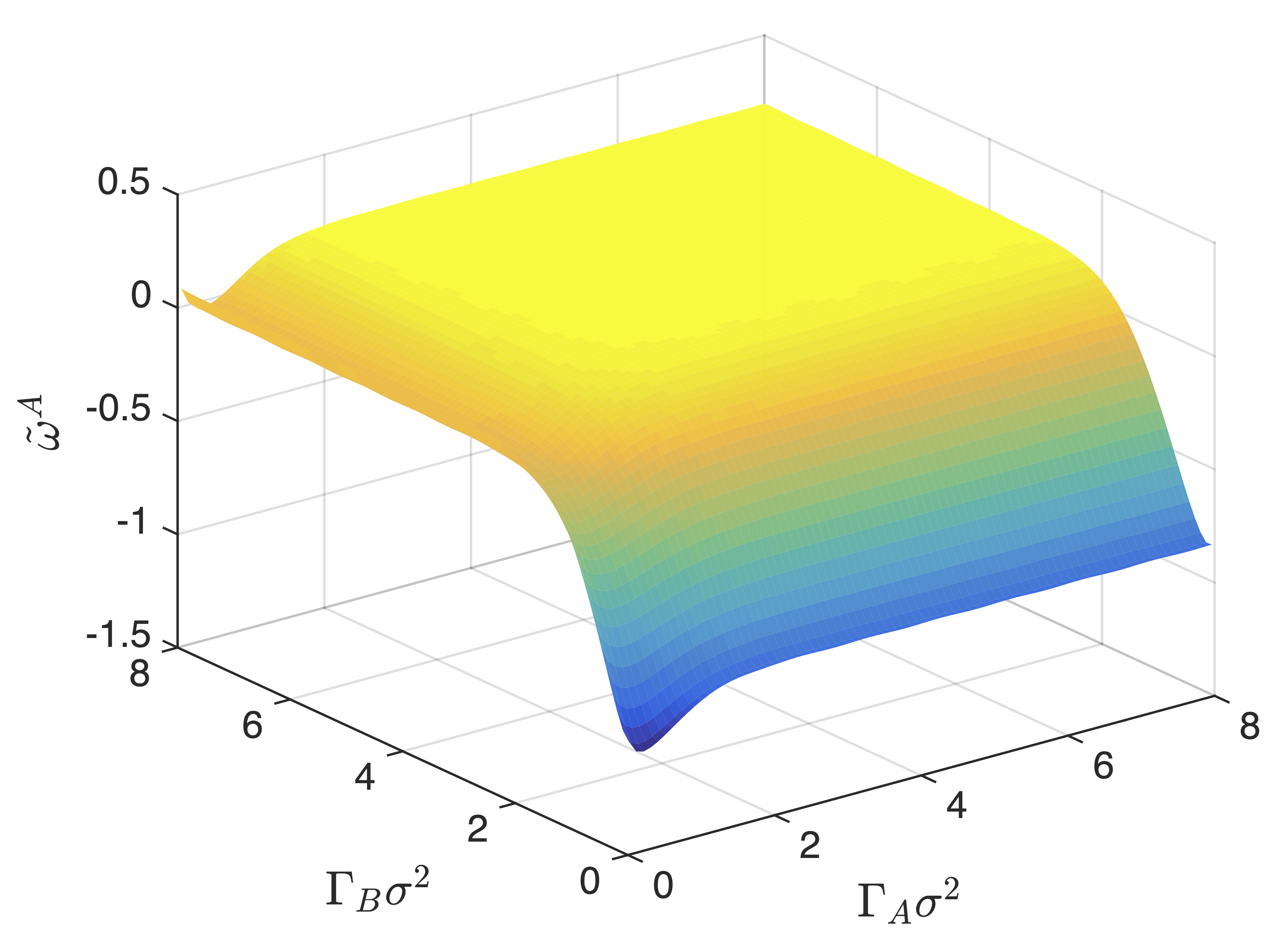}

(b)
\includegraphics[width=0.9\columnwidth]{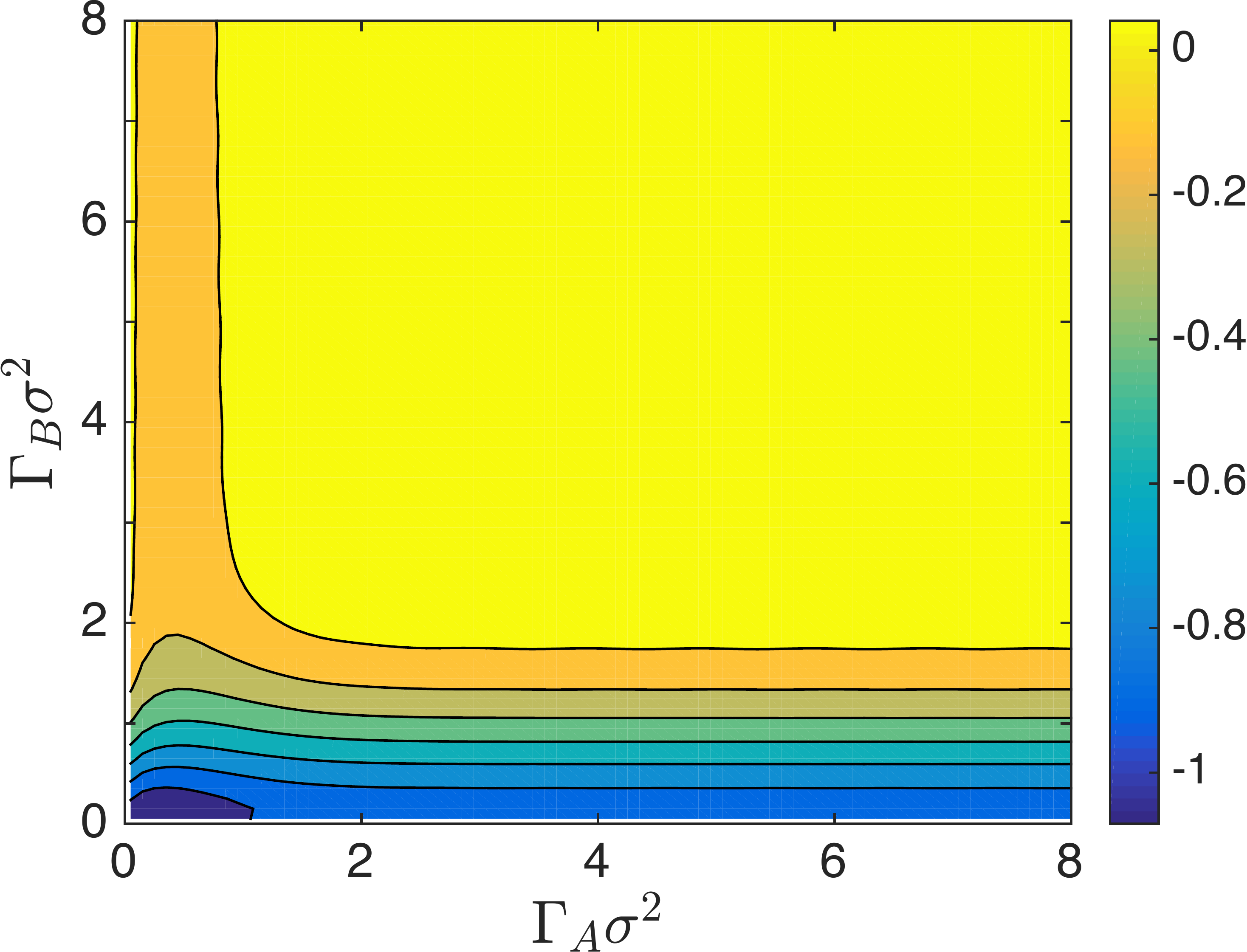}

\par\smallskip
\caption{The interface potential $\tilde{\omega}^{A}(\Gamma_{A}, \Gamma_{B})$ corresponding to the density profiles in Fig.~\ref{fig:de_mixing_case_down}, displayed as a surface in (a), and a contour plot (b). The stronger attraction between the wall and the species-$A$ particles leads to the wall favouring the liquid-$A$ phase being at the wall (with a small value of $\Gamma_B$), although since neither of the wall attraction strength parameters are large, the global minimum is for a small finite value of both adsorptions -- i.e.\ neither fluid wets the wall.}
\label{Omega_de_mixing_AB_2_5_down_general}
\end{figure}

Having discussed results for miscible liquids, we now move on to demonstrate what happens when the two species are immiscible.
Demixing typically occurs when the attraction between like-species particles is stronger than that between unlike-species. In such a situation, we can find two layers (films) of the different liquids at the surface, one on top of the other, in contact with the vapour. For the system to contain three different phases (vapour, liquid-$A$ and liquid-$B$), it must be at a state point in the vicinity of the triple point.

Fig.~\ref{fig:de_mixing_case_down} shows some typical examples of density profiles at a wall for a case where the two liquids are immiscible. The results here are for a mixture with $\varepsilon_{BB}=\varepsilon_{AA}$ and $\varepsilon_{AB}/\varepsilon_{AA}=0.75$. Since $\varepsilon_{AB}<\varepsilon_{AA} = \varepsilon_{BB}$, the particles of the two different species prefer to be next to their own kind and so demixing occurs. The results in Fig.~\ref{fig:de_mixing_case_down} are for temperature $k_{B}{T}/\varepsilon_{AA} = 0.67$ and chemical potentials $\mu_A/\varepsilon_{AA} = \mu_B/\varepsilon_{AA} = -4.5$. The wall parameters are chosen so that the wall attracts both particle species relatively weakly and therefore neither liquid-$A$ nor liquid-$B$ wets the wall. However, since the wall attracts the particles of species-$A$ more strongly than the particles of species-$B$, when we constrain a sizeable amount of each species to be at the wall, we find a film of the liquid-$A$ phase closest to the wall.

In Fig.~\ref{Omega_de_mixing_AB_2_5_down_general} we plot the interface potential $\tilde{\omega}^{A}(\Gamma_A,\Gamma_B)$ corresponding to the density profiles in Fig.~\ref{fig:de_mixing_case_down}. Since the lowest points on the surface $\tilde{\omega}^{A}(\Gamma_A,\Gamma_B)$ are when $\Gamma_B\approx0$, this shows that the system would rather have liquid-$A$ at the wall than liquid-$B$. The global minimum of the binding potential is not exactly at the origin, but the fact that it is at a finite value of both adsorptions, indicates that neither phase wets the wall. Note also that the binding potential is minimised when $\Gamma_{A} \neq \Gamma_{B}$, which is a further signature that the wall prefers one species over the other. Note that in the following section~\ref{sec:8} we plot this interface potential for fixed $\Gamma=\Gamma_A=\Gamma_B$, together with the solution branches to be discussed next and also examples of corresponding density profiles.

\begin{figure}[t]
\begin{center}
\includegraphics[width=0.9\columnwidth]{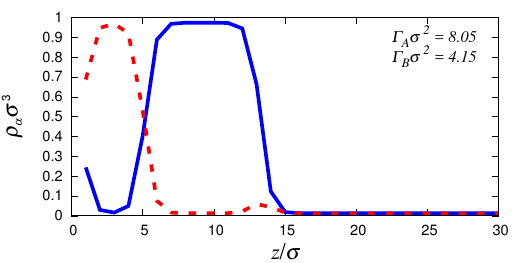}
\includegraphics[width=0.9\columnwidth]{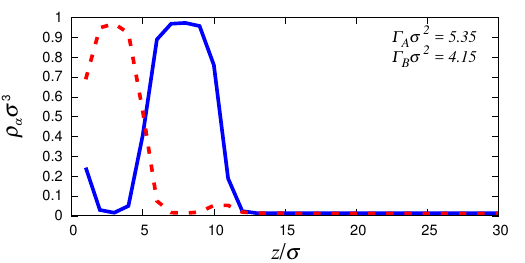}
\includegraphics[width=0.9\columnwidth]{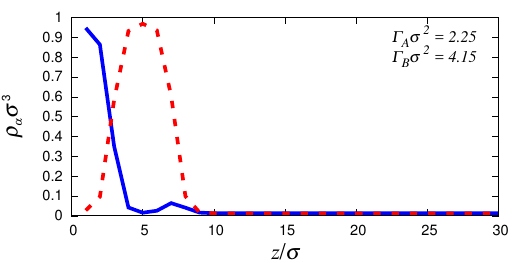}
\includegraphics[width=0.9\columnwidth]{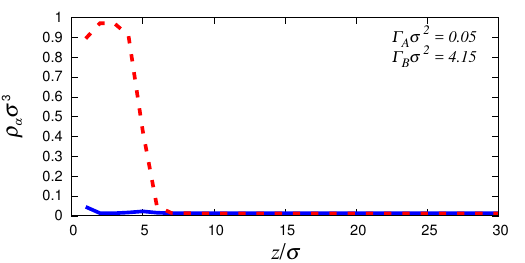}
\caption{Metastable density profiles for the same model parameters as the results in Fig.~\ref{fig:de_mixing_case_down}. Here the system has the liquid-$B$ film closest to the wall for the larger values of $\Gamma_A$ (in Fig.~\ref{fig:de_mixing_case_down} the liquid-$A$ film is closest to the wall). The solid (blue) lines are the density profiles of the species-$A$ and the dashed (red) lines are the profiles of the species-$B$.}
\label{fig:tran_1_fix_Gamma_b}
\end{center}
\end{figure}

\begin{figure}[t]

(a)
\includegraphics[width=0.9\columnwidth]{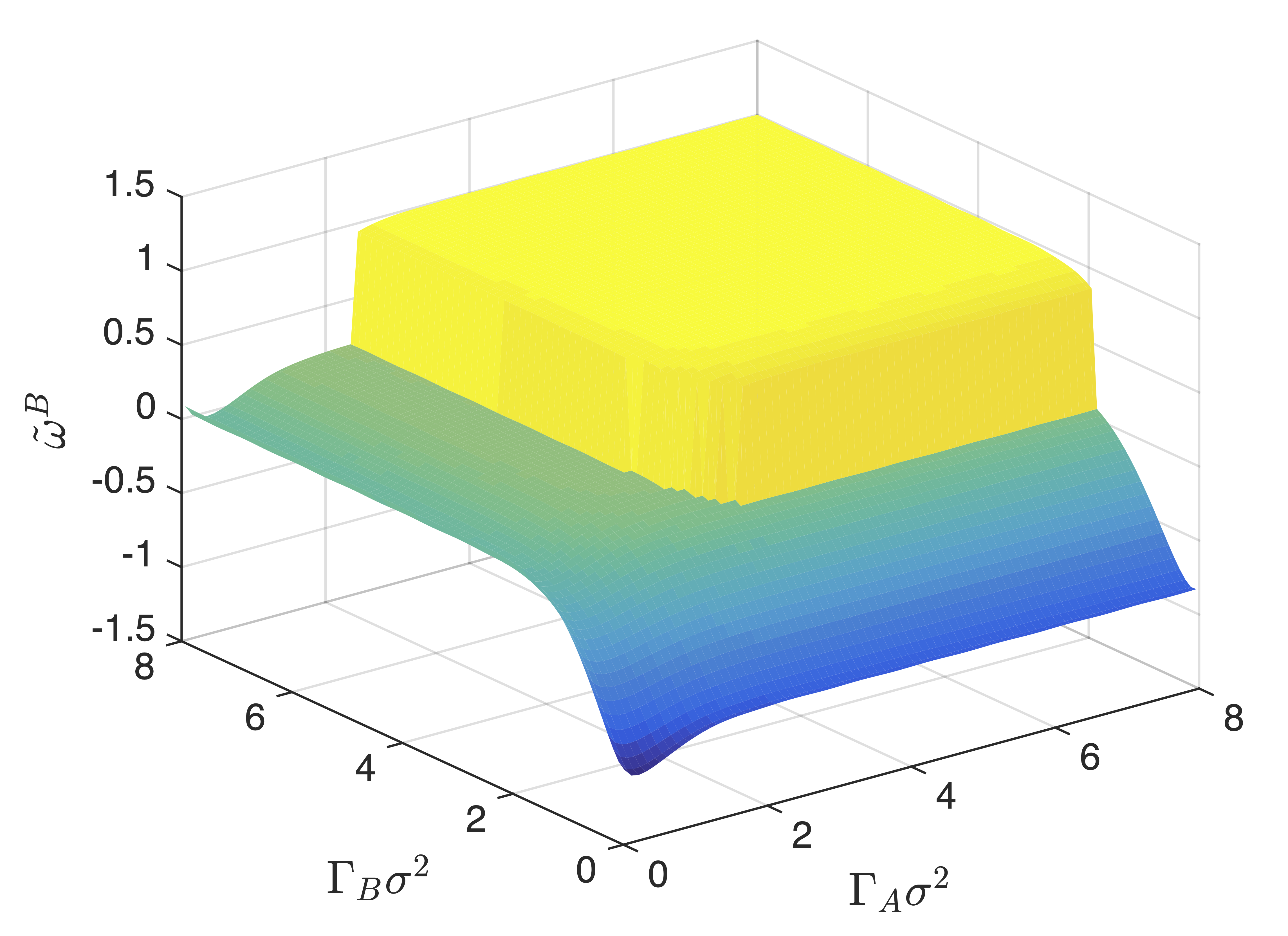}

(b)
\includegraphics[width=0.9\columnwidth]{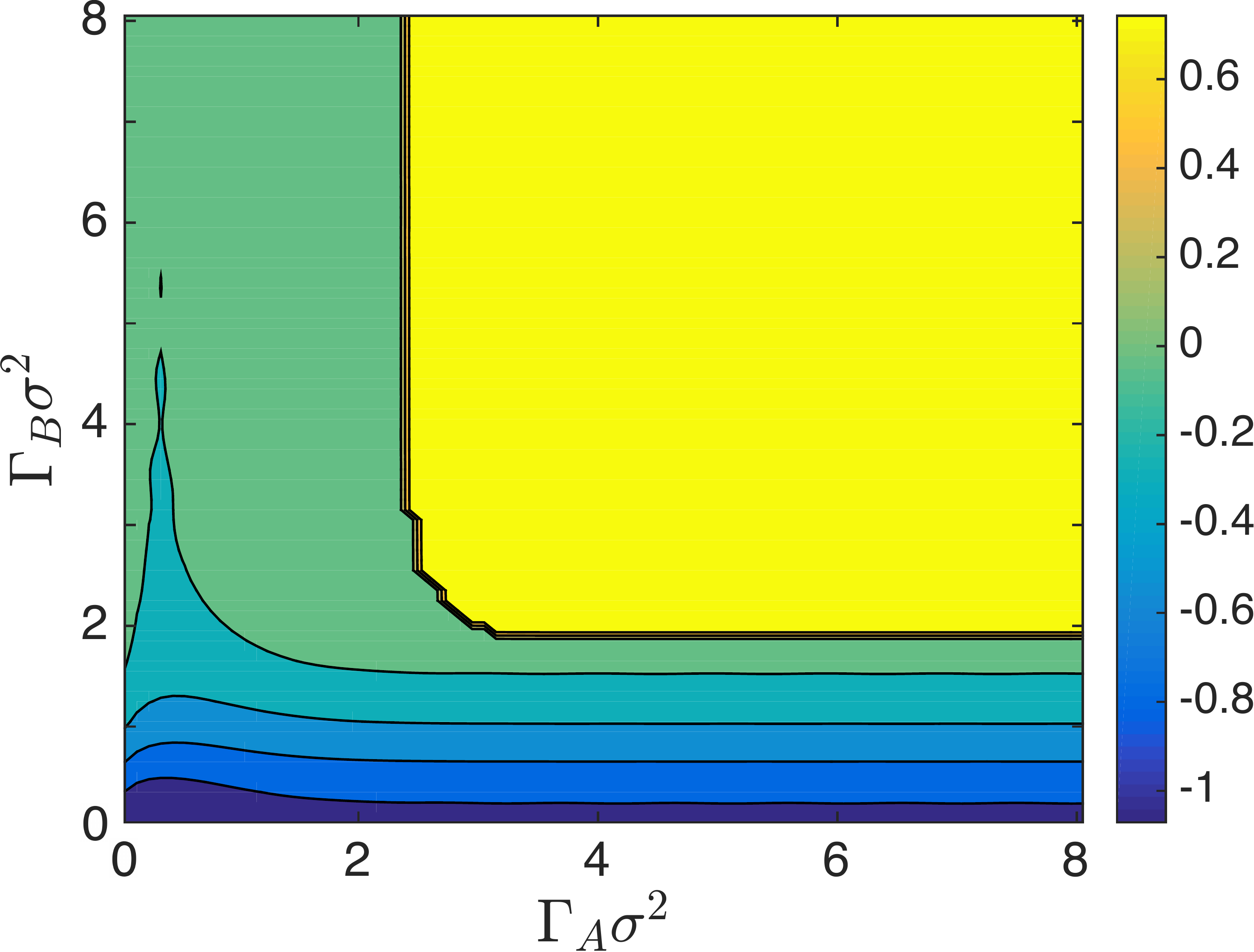}

\caption{The second branch of the interface potential $\tilde{\omega}^{B}(\Gamma_{A}, \Gamma_{B})$ (the yellow part), corresponding to the density profiles in Fig.~\ref{fig:tran_1_fix_Gamma_b}, displayed as a surface in panel~(a), and a contour plot in panel~(b). The parameters are as in Fig.~\ref{Omega_de_mixing_AB_2_5_down_general}. When one of the adsorptions $\Gamma_\alpha\sigma^2\lesssim 2$, the system falls off the upper (yellow) branch $\tilde{\omega}^{B}(\Gamma_{A}, \Gamma_{B})$, onto the lower branch $\tilde{\omega}^{A}(\Gamma_{A}, \Gamma_{B})$ that is in Fig.~\ref{Omega_de_mixing_AB_2_5_down_general} also displayed for the full range of the adsorptions.}
\label{Omega_de_mixing_AB_2_5_general}
\end{figure}

Although the wall prefers to be in contact with liquid-$A$ rather than liquid-$B$, it is possible to initiate the system with a film of liquid-$B$ closest to the wall, with a film of liquid-$A$ above it. In Fig.~\ref{fig:tran_1_fix_Gamma_b} we display equilibrated density profiles corresponding to this situation. The profiles in this figure are for exactly the same parameters as used in the previous case in Fig.~\ref{fig:de_mixing_case_down}, but they correspond to a separate branch of metastable solutions to the equations (local minima of the free energy, but not the global minimum) and therefore we find a second interface potential $\tilde{\omega}^{B}$, applicable when the liquid-$B$ phase is closest to the wall and liquid-$A$ is above it. The density profiles in Fig.~\ref{fig:tran_1_fix_Gamma_b} reflect the fact that the attraction between species-$A$ particles and the wall is stronger than between species-$B$ particles and the wall, i.e.\ we see that the density profiles of species-$A$ have a maximum close to the wall. This branch of solutions does not exist for all values of $(\Gamma_A,\Gamma_B)$. If either of the adsorptions values become small, $\Gamma_\alpha\sigma^2\lesssim2.5$, then the system rearranges itself, decreasing the free energy, to be on the $\tilde{\omega}^A$ branch of solutions. In Fig.~\ref{Omega_de_mixing_AB_2_5_general} we display both ${\tilde{\omega}}^B$ and ${\tilde{\omega}}^A$. The upper (yellow) surface corresponds to ${\tilde{\omega}}^B$ and the lower surface corresponds to ${\tilde{\omega}}^A$. A similar multi-stability is discussed in Ref.~\cite{ThMF2007pf}, where a Cahn-Hilliard model for a binary mixture between walls with energetic bias is analyzed. Ref.~\cite{ThMF2007pf} also presents bifurcation diagrams, showing the relations and connections of various branches of stable, metastable and unstable layered configurations. With the methods used here, we are solely able to track stable solution branches, but comparison with Ref.~\cite{ThMF2007pf} gives hints as to how the solution branches found here may possible be connected. \red{Note also that the somewhat ragged edge to the upper (yellow) branch in Fig.~\ref{Omega_de_mixing_AB_2_5_general} is due to the finite size steps taken in scanning the $(\Gamma_A,\Gamma_B)$ parameter space. With a smaller step size, this raggedness would be reduced. This remark also applies to Fig.~\ref{de-mixing case for another sheet} below.}

\begin{figure}[t!]
\begin{center}
\includegraphics[width=0.9\columnwidth]{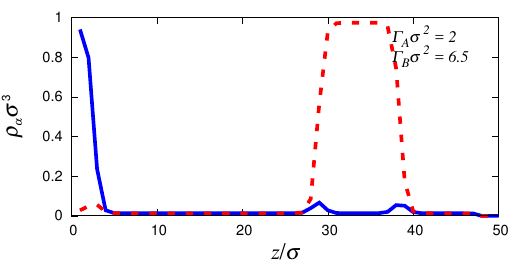}
\includegraphics[width=0.9\columnwidth]{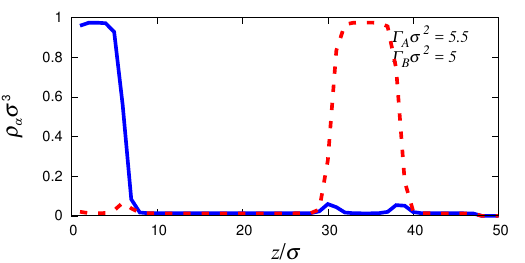}
\includegraphics[width=0.9\columnwidth]{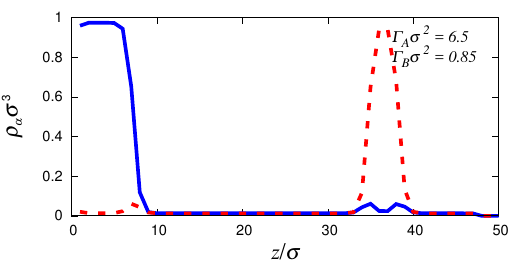}
\includegraphics[width=0.9\columnwidth]{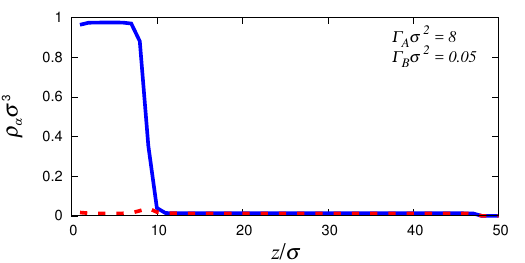}

\caption{Metastable density profiles for the same model parameters as the results in Fig.~\ref{fig:de_mixing_case_down}, except here find the system with a film of vapour between the two liquid layers. The solid (blue) line is for the species-$A$ particles and the dashed (red) line is for the species-$B$ particles.}
\label{fig:demixing_case_gas}
\end{center}
\end{figure}

\begin{figure}[t]
(a)
\includegraphics[width=0.9\columnwidth]{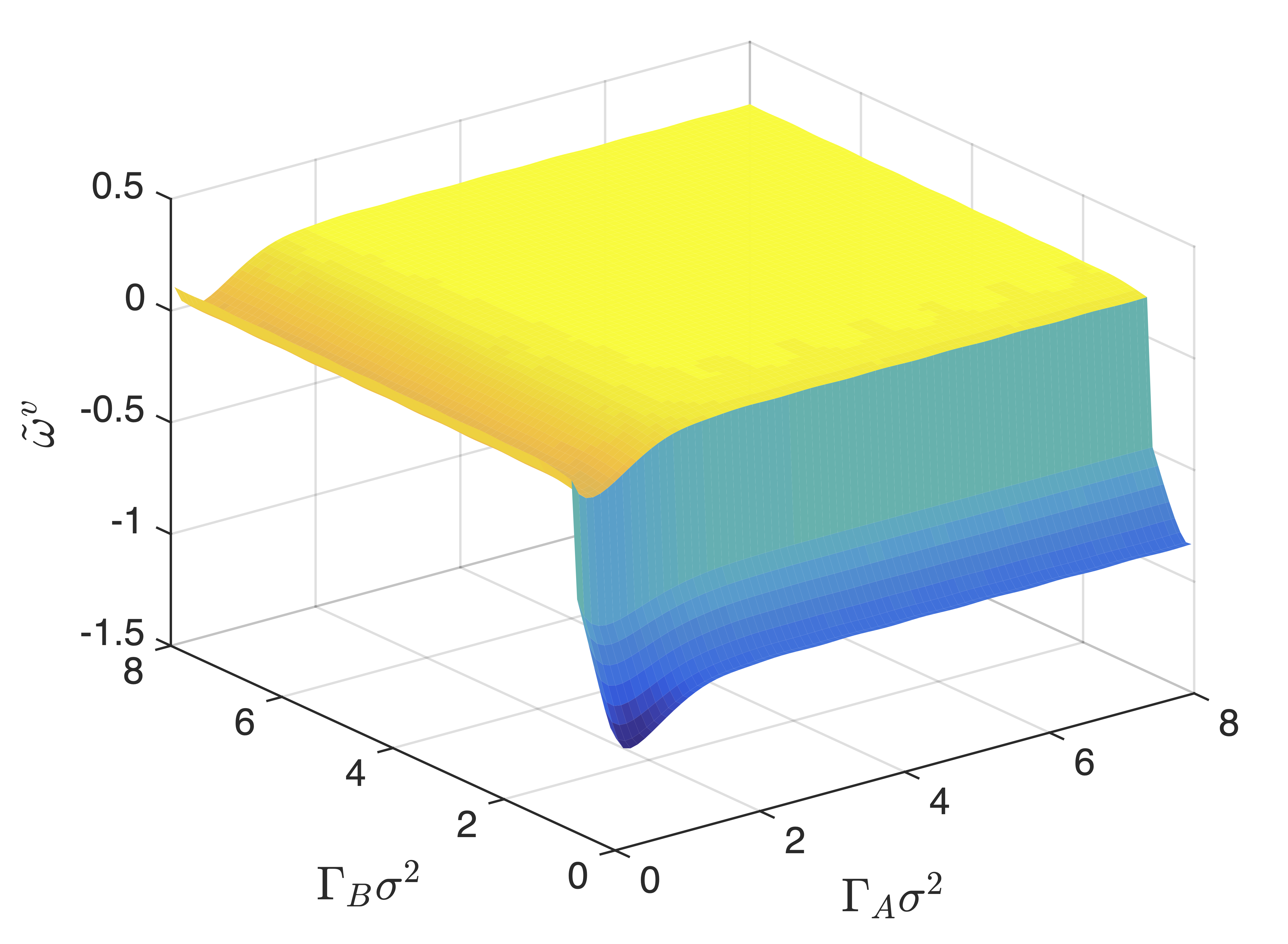}

(b)
\includegraphics[width=0.9\columnwidth]{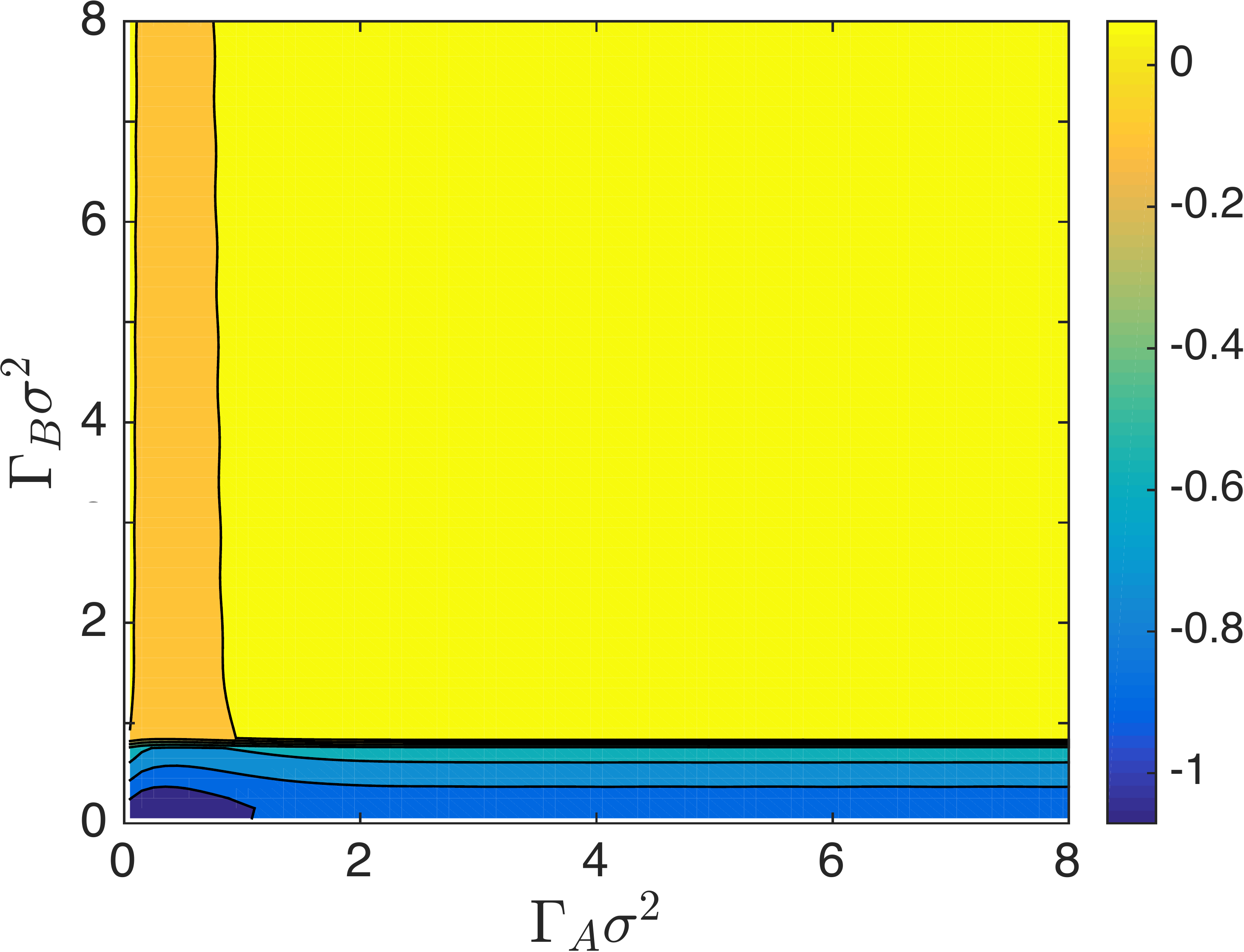}

\caption{The interface potential $\tilde{\omega}^{v}(\Gamma_{A}, \Gamma_{B})$ corresponding to the density profiles in Fig.~\ref{fig:demixing_case_gas}, displayed as a surface in (a), and a contour plot in (b). These are for exactly the same parameters as the results in Fig.~\ref{Omega_de_mixing_AB_2_5_down_general}. When the species-$B$ adsorption $\Gamma_B\sigma^2\lesssim 1$, the system falls off the upper (yellow) interface potential $\tilde{\omega}^{v}(\Gamma_{A}, \Gamma_{B})$, onto the lower (green-blue) $\tilde{\omega}^{A}(\Gamma_{A}, \Gamma_{B})$ branch that is also displayed in Fig.~\ref{Omega_de_mixing_AB_2_5_down_general}.} 
\label{Omega_de_mixing_AB_2_5_general_gas}
\end{figure}

The two different branches of solutions to the equations described above are not the only possible solutions that we find. Additionally, we display in Fig.~\ref{fig:demixing_case_gas} density profiles for exactly the same parameters used in the previous cases in this section. These density profiles correspond to a case where a layer of vapour has intruded between the two layers of the immiscible liquids. For example, when $\Gamma_A\sigma^2 = 2$ and $\Gamma_B\sigma^2 = 6.5$, we observe a thin layer of the liquid-$A$ phase adsorbed at the wall and a thicker film of the liquid-$B$ phase far away from the wall with a layer of the vapour between the two. We should emphasise that such a configuration is not the global minimum of the free energy. Moreover, a factor that may help stabilise such a configuration is the fact that the system is discretised on a lattice. Oscillations (i.e.\ multiple minima) in the binding potential for lattice-gas models were discussed previously for one-component systems -- see Ref.~\cite{hughes2015liquid}. In Fig.~\ref{Omega_de_mixing_AB_2_5_general_gas} we display the interface potential of $\tilde{\omega}^v$ corresponding to these configurations. When $\Gamma_B$ becomes small enough this metastable configuration branch of solutions ceases to exist and the system falls down onto the $\tilde{\omega}^A$ branch of solutions.

Another metastable branch of solutions that we were able to find for this set of parameters includes configurations (not displayed) corresponding to the liquid-$B$ phase being at the wall, above which is a film of the vapour, and then above that a film of liquid-$A$. The corresponding interface potential $\tilde{\omega}^B$ is displayed in Fig.~\ref{de-mixing case for another sheet}. As for some of the previous metastable configurations, we believe the fact that we are dealing with a lattice model where liquid-vapour interfaces can get pinned to the underlying grid plays a role in stabilising these. However, there are continuum systems where the liquid-liquid interfacial profiles exhibit oscillations (see e.g.\ Ref.~\cite{archer2001binary, hughes2017}) and the weak pinning of the interfaces that we observe here due to the lattice could also in principle occur in such systems. Binding potentials with oscillatory decay have been observed, also leading to the pinning of interfaces \cite{hughes2017, yin2017films}. In the following section we present further details on this system by displaying slices through the various interface potentials together with some typical density profiles.

\begin{figure}[t]
\centering
(a)
\includegraphics[width=0.9\columnwidth]{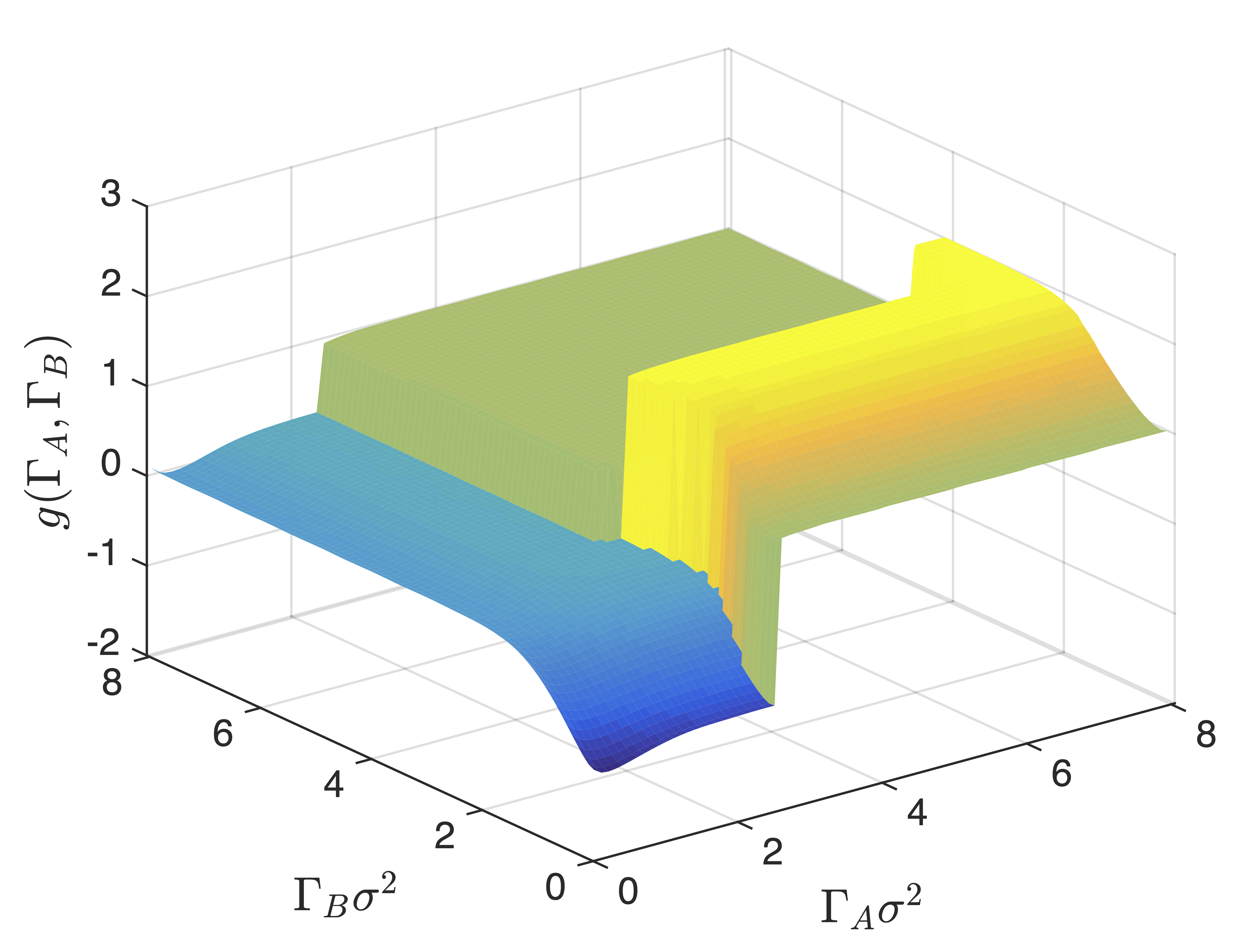}

(b)
\includegraphics[width=0.9\columnwidth]{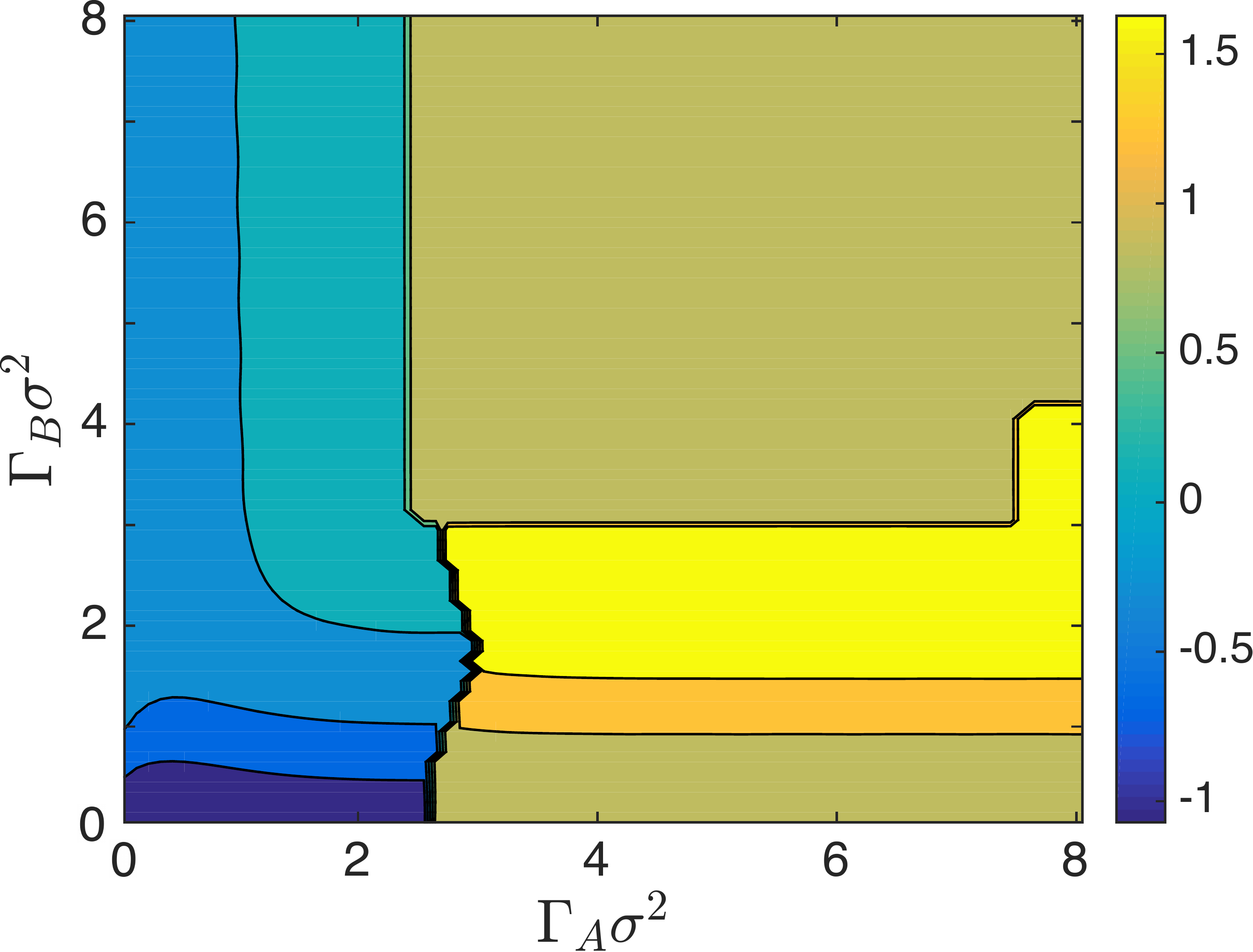}

\caption{Portions of three different interface potentials $\tilde{\omega}^{B}(\Gamma_{A}, \Gamma_{B})$ corresponding to a case where the system was initiated with a film of liquid-$B$ at the wall, above this a film of vapour, then a film of liquid-$A$ phase, before finally arriving at the bulk vapour. Such configurations are rather fragile metastable states that have a much larger energy than the globally stable state, with a relatively low energy barrier to be overcome to leave the metastable state, which is why only portions of the surface were obtained. These are for exactly the same parameters as the results in Fig.~\ref{Omega_de_mixing_AB_2_5_down_general}.}
\label{de-mixing case for another sheet}
\end{figure}
\section{Binding potentials and profiles for $\Gamma_A=\Gamma_B$}
\label{sec:8}

In this section we present further details and additional results for the demixing liquid system discussed in the previous section. In particular, we plot the interface potentials along the line $\Gamma_{A} = \Gamma_{B}=\Gamma$, where the two adsorptions are equal, together with some of the corresponding density profiles.

\begin{figure*}[t!]

\includegraphics[width=0.99\textwidth]{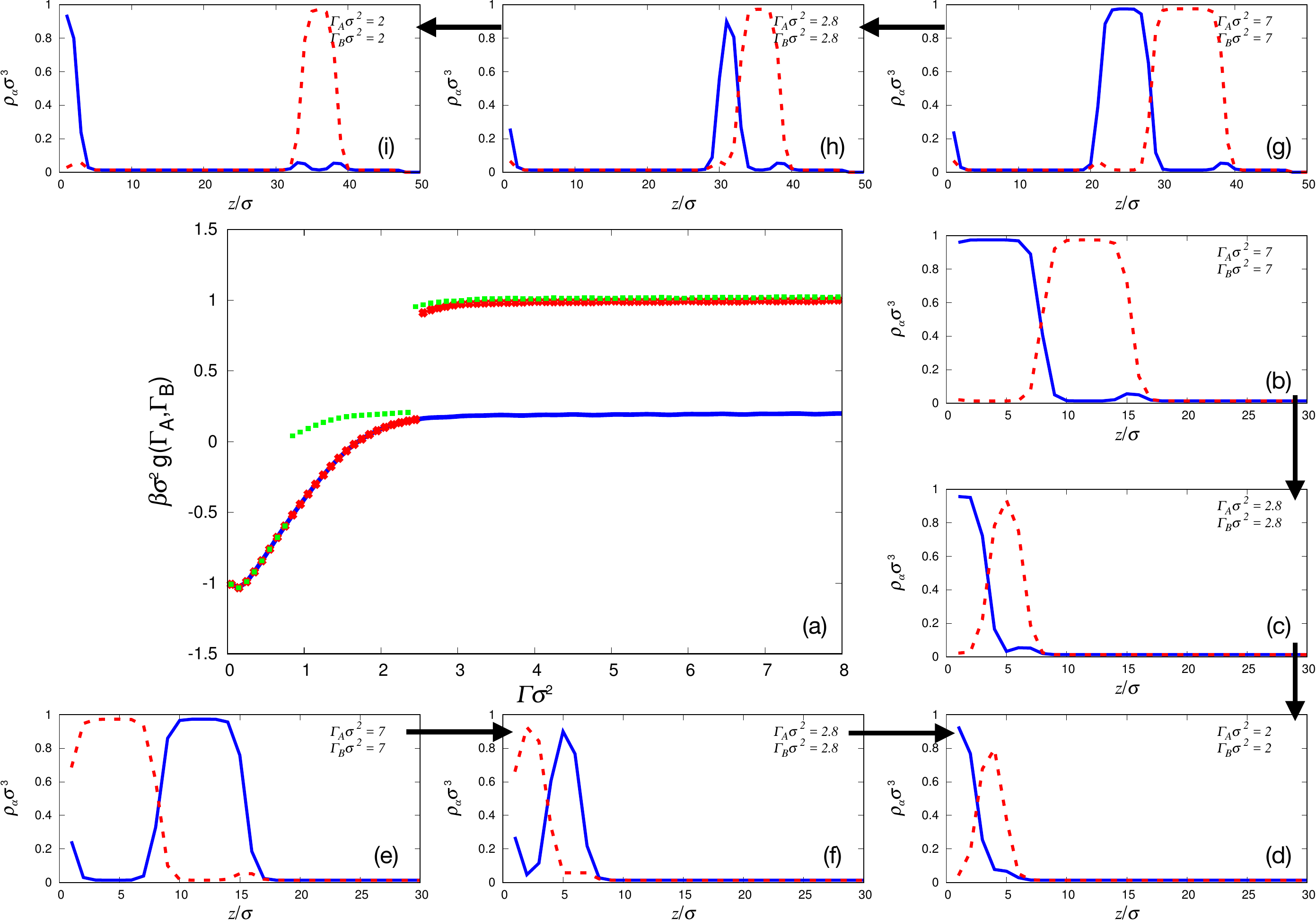}

\caption{In (a) we display the interface potentials $\tilde{\omega}^A$ (blue line), $\tilde{\omega}^B$ (red points) and $\tilde{\omega}^v$ (green points) as functions of $\Gamma = \Gamma_A = \Gamma_B$ (i.e.\ both adsorptions are equal), calculated for decreasing $\Gamma$. The system parameters are the same as the results in Figs.~\ref{fig:de_mixing_case_down}--\ref{de-mixing case for another sheet}. The blue line is a slice through the $\tilde{\omega}^A$ surface displayed in Fig.~\ref{Omega_de_mixing_AB_2_5_down_general}, which corresponds to the lowest free energy case, with a film of liquid-$A$ closest to the wall. Some examples of density profiles along this branch are displayed in panels (b), (c) and (d) (down the right), with the values of $\Gamma_A=\Gamma_B$ given in the keys. The red points correspond to a different branch of solutions (displayed in Fig~\ref{Omega_de_mixing_AB_2_5_general}), corresponding to liquid-$B$ being closest to the wall. Examples of density profiles along this branch are displayed in panels (e), (f) and (d) (along the bottom). Notice that at $\Gamma\sigma^2\approx2.5$ this branch ends and the system falls onto the $\tilde{\omega}^A$ branch (blue line). The green points correspond to the $\tilde{\omega}^v$ branch of solutions where there is a film of the vapour phase intruding between the two films of liquid and the wall. Examples of density profiles along this branch are displayed along the top in panels (g), (h) and (i). Notice that as $\Gamma$ is decreased the system first falls onto a branch of solutions where liquid-$A$ is at the wall, whilst liquid-$B$ remains above the trapped vapour film. Then, subsequently, it falls onto the $\tilde{\omega}^A$ branch of solutions.}
\label{fig:different_Gamma_2_5}
\end{figure*}

In Fig.~\ref{fig:different_Gamma_2_5}(a) we display three different interface potential curves calculated for decreasing $\Gamma$. These results are for the immiscible liquid mixture considered in the previous section and for the same state point. The blue curve is $\tilde{\omega}^A$, which has a film of liquid-$A$ closest to the wall, with a film of liquid-$B$ between liquid-$A$ and the bulk vapour. This is the lowest free energy configuration for any given value of $\Gamma$. Examples of density profiles along this branch of solutions are displayed in panels (b), (e) and (h) of Fig.~\ref{fig:different_Gamma_2_5}.

Also in Fig.~\ref{fig:different_Gamma_2_5}(a) the red points for larger $\Gamma$ lie on the branch $\tilde{\omega}^B$, which corresponds to the two liquid films at the wall being the other way round, i.e.\ with the film of liquid-$B$ at the wall. However, at $\Gamma\sigma^2\approx2.5$ this branch ends and the system falls down onto the lower free energy $\tilde{\omega}^A$ branch. Examples of density profiles from along this sequence are displayed in panels (c), (f) and (i) of Fig.~\ref{fig:different_Gamma_2_5}.

Finally, in Fig.~\ref{fig:different_Gamma_2_5}(a) the green points start for larger $\Gamma$ on the branch $\tilde{\omega}^v$, which corresponds to having a film of the vapour intruding between the two liquid films and the wall. At $\Gamma\sigma^2\approx2.5$ this branch ends and the system falls down onto a lower (but still metastable) free energy branch where liquid-$A$ is at the wall, but liquid-$B$ remains above the vapour layer. Then subsequently at $\Gamma\sigma^2\approx1$ this second branch ends and the system falls down onto the lowest free energy $\tilde{\omega}^A$ branch. Examples of density profiles from along this sequence are displayed in panels (d), (g) and (j) of Fig.~\ref{fig:different_Gamma_2_5}. Note that this branch of solutions corresponding to a vapour layer between the liquids and the wall is not the same as the branch of solutions investigated in Figs.~\ref{fig:demixing_case_gas} and \ref{Omega_de_mixing_AB_2_5_general_gas}, which correspond to a layer of vapour intruding between the liquid-$A$ and liquid-$B$ films, with the liquid-$B$ layer at the wall.

\section{Density profiles of droplets at walls}
\label{sec:9}

\begin{figure}[t!]
\centering
\includegraphics[width=0.95\columnwidth]{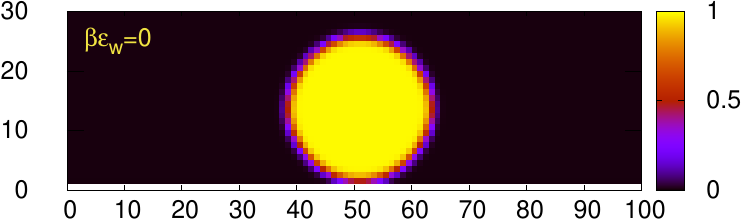}

\includegraphics[width=0.95\columnwidth]{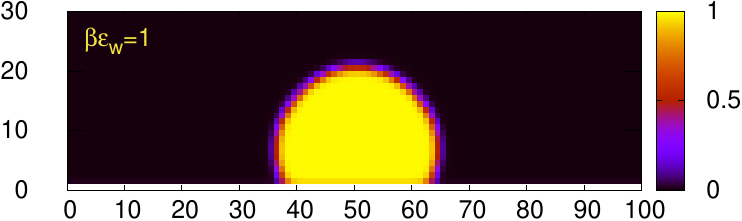}

\includegraphics[width=0.95\columnwidth]{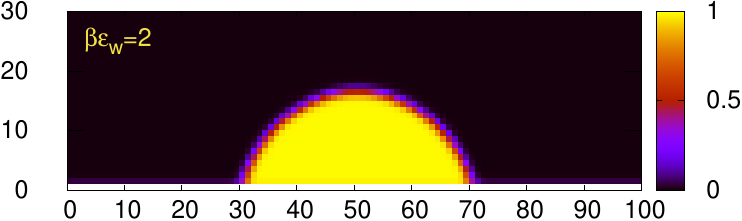}

\includegraphics[width=0.95\columnwidth]{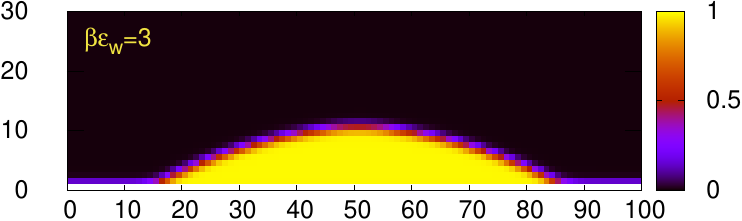}

\caption{A series of 2D droplet profiles for a symmetric miscible binary mixture with $\beta\varepsilon_{AA} = 1.5$, $\varepsilon_{BB}=\varepsilon_{AA}$ and $\varepsilon_{AB}/\varepsilon_{AA} = 1.2$, at walls of varying attraction strength that is identical for the two species. The computational domain is of size $L_x=100$ in the horizontal direction and $L_z=33$ in the vertical direction. We assume that the density is invariant in the $y$-direction. We assume the upper wall (at $z=33$) is purely repulsive. The total volumes of liquid-$A$ and liquid-$B$ are equal, with a total of $n_A=n_B=245$ particles per unit area in the system. Due to the symmetry, the density difference $\rho_\mathbf{i}^A-\rho_\mathbf{i}^B=0$ for all $\mathbf{i}$. Therefore, here we solely display the total density $\rho_\mathbf{i}^A+\rho_\mathbf{i}^B$. The droplets are deposited on various solid substrates with attraction strength parameters $\beta\varepsilon_{wA} = \beta\varepsilon_{wB} = \beta\varepsilon_w=0$, 1, 2 and 3.}
\label{fig:21}
\end{figure}

\begin{figure*}[t!]
\centering
\includegraphics[width=0.95\columnwidth]{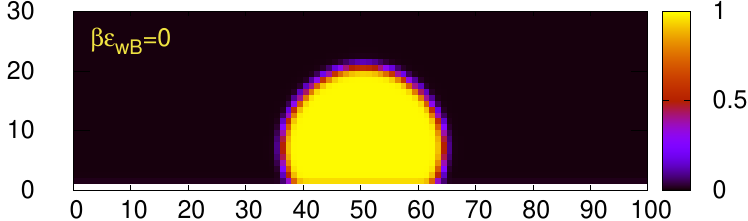}
\includegraphics[width=0.95\columnwidth]{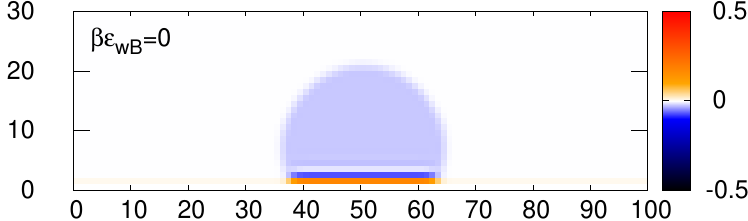}

\includegraphics[width=0.95\columnwidth]{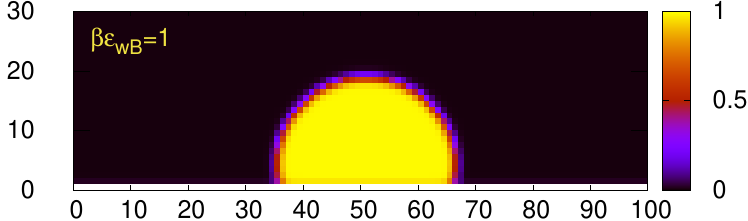}
\includegraphics[width=0.95\columnwidth]{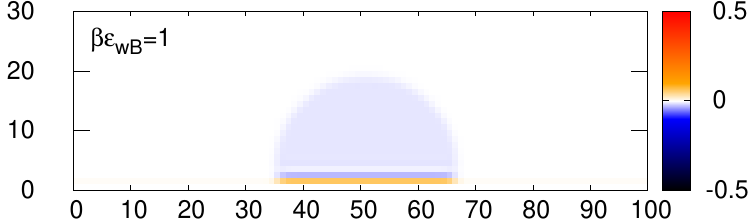}

\includegraphics[width=0.95\columnwidth]{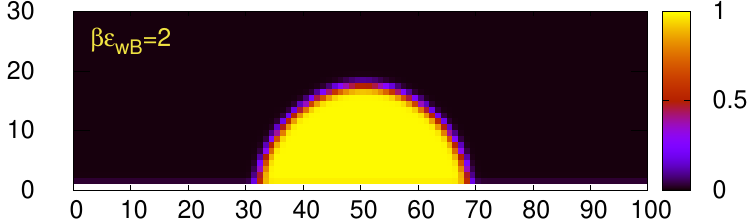}
\includegraphics[width=0.95\columnwidth]{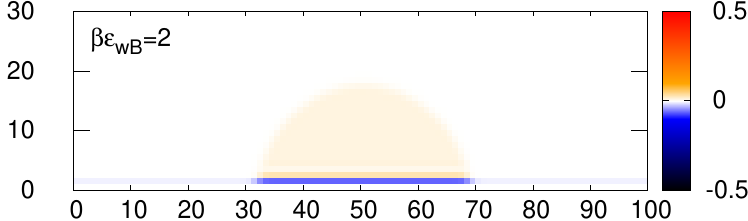}

\includegraphics[width=0.95\columnwidth]{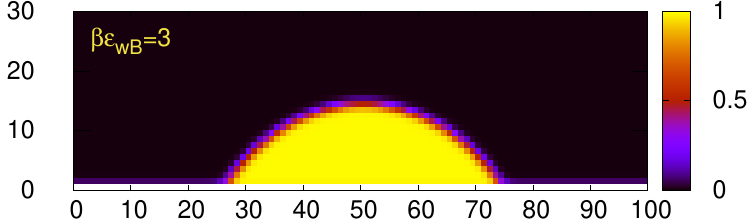}
\includegraphics[width=0.95\columnwidth]{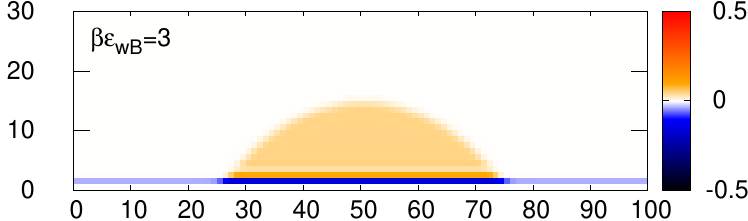}

\caption{A series of 2D droplet profiles for the same symmetric miscible binary mixture considered in Fig.~\ref{fig:22}, with the same equal total volumes of liquid-$A$ and liquid-$B$ but for wall interactions that are  different for the two species. In particular, we fix the wall-species-$A$ attraction strength to be $\beta\varepsilon_{wA}=1.5$, while we vary the wall-species-$B$ attraction strength, choosing the values $\beta\varepsilon_{wB}=0, 1, 2$ and $3$. On the left, we display the total fluid density $\rho_\mathbf{i}^A+\rho_\mathbf{i}^B$, while on the right we display the corresponding density difference $\rho_\mathbf{i}^A-\rho_\mathbf{i}^B$.}
\label{fig:22}
\end{figure*}

\begin{figure}[t!]
\centering
\includegraphics[width=0.95\columnwidth]{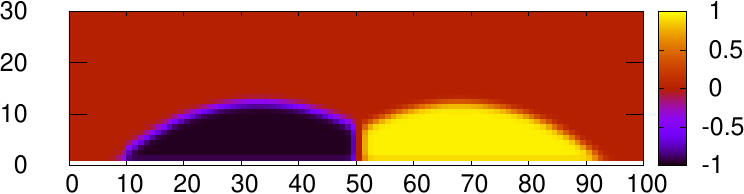}

\includegraphics[width=0.95\columnwidth]{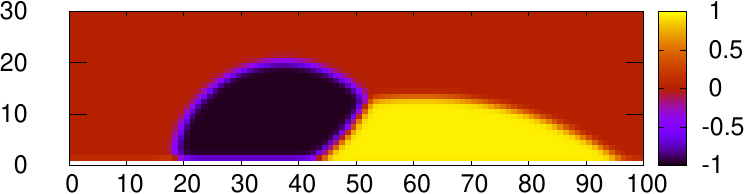}

\includegraphics[width=0.95\columnwidth]{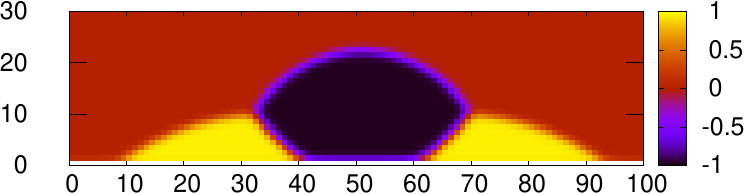}

\caption {Droplet profiles for two immiscible liquids deposited on a surface, plotting the density difference $\rho_\mathbf{i}^A-\rho_\mathbf{i}^B$. In all three cases the interaction parameters are $\varepsilon_{AA}/k_BT = 1.5$, $\varepsilon_{AB}/\varepsilon_{AA} = 0.5$ and $\varepsilon_{BB}=\varepsilon_{AA}$. For the top profile the wall interaction strengths are equal, with $\varepsilon_{wA}/k_BT = \varepsilon_{wB}/k_BT = 1.5$, while for the lower two cases, the wall prefers species-$A$, with $\varepsilon_{wA}/k_BT = 2.5$ and $\varepsilon_{wB}/k_BT = 1.5$. For the top case, the volume of the two fluids are equal, with a total of $n_A=n_B=420$ particles per unit area in the system. For the middle case, $n_A=n_B=525$, while for the bottom plot, $n_A=420$ and $n_B=595$.}
\label{fig:23} 
\end{figure}

Recall the results in Figs.~\ref{fig:7} and \ref{fig:9} displaying density profiles for a symmetric miscible binary mixture in contact with a planar wall that attracts the two different species equally. In the case in Fig.~\ref{fig:7}, the fluid does not wet the wall, since the wall-attraction strength $\varepsilon_{wA}=\varepsilon_{wB}$ is not large enough. This can also be seen from inspecting the corresponding binding potential, displayed in Fig.~\ref{fig:10}. In contrast, when the wall attraction is much stronger (e.g.\ the case in Fig.~\ref{fig:9}), the fluid wets the wall. This is also indicated by the binding potential in Fig.~\ref{fig:12}. These density profiles were calculated assuming that the liquid density only varies in the $z$-direction, the direction perpendicular to the wall. However, for partially wetting systems, one should expect a uniform thickness film of liquid deposited on the surface to dewet, to form droplets with a finite contact angle. This is indeed what we observe: In Fig.~\ref{fig:21} we display droplets of the symmetric miscible binary mixture considered already in Figs.~\ref{fig:7}--\ref{fig:12} with $\varepsilon_{AA}=\varepsilon_{BB}$ and $\varepsilon_{AB}=1.2\varepsilon_{AA}$. We observe that as the wall attraction strength that is identical for the two species ($\varepsilon_{wA}=\varepsilon_{wB}=\varepsilon_{w}$) is increased from zero, the contact angle of the droplets decreases. Due to the complete symmetry between components, the density profiles of the two different species are identical and so we display in Fig.~\ref{fig:21} just the total density $\rho_\mathbf{i}^A+\rho_\mathbf{i}^B$. These profiles are calculated using the same procedure as used to calculate the 1D profiles in e.g.\ Fig.~\ref{fig:7}, except here we make the system wide enough in one direction parallel to the surface that density variations in this direction become possible -- see Refs.~\cite{hughes2014introduction, hughes2015liquid, hughes2017} for discussion and examples of this for one-component fluids. We still constrain the adsorption at the surface, though now for these systems varying in 2D, a better measure for the amount of material in the system is the number of particles per unit area, $n_\alpha=\sum_\text{2D area}\rho_\mathbf{i}^\alpha$. The summation here denotes a summation performed over the $(x,z)$-plane, while we assume that the density is invariant in the $y$-direction (into the plane of the figure). Thus, the droplet profiles displayed here actually correspond to cross-sections through ridge-shaped droplets. For the droplets in Fig.~\ref{fig:21}, we fix $n_A=n_B=245$ to be the same in all cases. We observe that as $\varepsilon_w$ is increased, the contact angle decreases until at $\beta\varepsilon_w\approx3.3$ there is a wetting transition, where the contact angle $\theta\to0$.

For the density profiles in Fig.~\ref{fig:21}, we vary the wall attraction strength while also keeping it identical for the two different species of particles. In contrast, in Fig.~\ref{fig:22} we keep the wall-species-$A$ attraction strength fixed at $\beta\varepsilon_{wA}=1.5$, while we vary the wall-species-$B$ attraction strength, $\varepsilon_{wB}$. We find that increasing $\varepsilon_{wB}$ leads to the droplet spreading further over the surface and thus to a decrease in the contact angle, with the wetting transition occurring at $\beta\varepsilon_{wB}\approx4.2$ (for this fixed equal total average densities of the two species). Since the interaction with the wall is no longer symmetric, the density profiles of the two different species are no longer identical. In Fig.~\ref{fig:22} we plot on the left hand side the total density $\rho_\mathbf{i}^A+\rho_\mathbf{i}^B$ for varying $\varepsilon_{wB}$, while on the right hand side we plot the difference $\rho_\mathbf{i}^A-\rho_\mathbf{i}^B$. This allows to see that when the wall attracts species-$A$ more strongly than species-$B$ (cases at the top of Fig.~\ref{fig:22}), then the local density of species-$A$ in the layer directly in contact with the wall is higher than that of species-$B$ (a similar case is displayed in Fig.~\ref{fig:8}). Interestingly, in the layer directly above this, it is the other way round. Moreover, the increased density of species-$A$ at the wall leads to the overall density of species-$A$ in the bulk of the droplet being slightly lower than the density of the $B$ particles. As we then increase the attraction between the wall and the species-$B$ particles, the roles are reversed, i.e.\ when the wall attracts species-$B$ more strongly than species-$A$ (i.e.\ cases at the bottom of Fig.~\ref{fig:22}), then the local density of species-$B$ in the layer right at the surface of the wall is higher than that of species-$A$ particles. Correspondingly, in the layer directly above this, the density of species-$A$ is higher than that of species-$B$ and also in the bulk of the droplet.

Having considered examples of droplets of miscible liquids on surfaces, we display in Fig.~\ref{fig:23} a few examples of immiscible liquids on surfaces. Here we plot the density difference $\rho_\mathbf{i}^A-\rho_\mathbf{i}^B$, with the colour bar chosen so that the bulk vapour is in red, droplets of liquid-$A$ appears in yellow, while droplets of liquid-$B$ appears black. In the top panel, the attraction strengths between the wall and the two species are identical, so due to the symmetry in the system, the two droplets are mirror-identical to one another. In contrast, for the droplets in the lower two plots the wall prefers species-$A$, so the yellow droplets have a smaller contact angle with the wall than the black liquid-$B$ droplets. Here, the middle panel shows an asymmetric compound drop while the lower configuration is fully left-right symmetric. These droplet configurations are just a few examples of possible arrangements of the liquids. Depending on the volumes of each of the liquids and the size of the system, various different (meta)stable configurations are possible. Such compound droplets and their transitions are also studied for macroscopic and mesoscopic sharp-interface models \cite{PBMT2005jcp, NTGD2012sm, Dieckmann2022Munster}, as well as with a diffuse interface approach \cite{BSNB2014l}. In Ref.~\cite{BSNB2014l, Dieckmann2022Munster} both, symmetric and asymmetric states, are found in contrast to Ref.~\cite{MaAP2002jfm} where it is argued that asymmetric compound drops cannot exist and to Ref.~\cite{NTGD2012sm} where they seem to be attributed to the action of gravity. We note also that not only is the statics of binary liquid droplets on solid substrates varied and interesting; the nonequilibrium dynamics is also complex and striking -- see e.g.\ Refs.~\cite{pototsky2004alternative,PBMT2005jcp,PBMT2006el,BMBK2011epje, KHFR2014l, KaRi2014jfm, KaRi2012prl, KaRi2010l,IDSS2017l,edwards2018density,DGGR2021l,NeSi2021prf}. We defer comprehensive study of possible arrangements of droplets on surfaces to future work.

\section{Concluding remarks}\label{sec:conc}

We have developed a theory for calculating binding potentials as functions of the adsorptions, for liquid mixtures at interfaces. We have shown that the method can be applied to both miscible and immiscible liquids. Our approach is based on DFT and so also yields the density profiles of the two liquids at the surfaces, alongside other thermodynamic quantities. Here, we have applied the method to a simple binary lattice-gas model to demonstrate its effectiveness for determining crucial quantities for understanding the interfacial wetting phase behaviour of binary mixtures of liquids. However, the method can in future be applied with any other more sophisticated and/or accurate DFT and for other types of binary-mixture systems.

The method allows one to go from the microscopic pair interaction potential parameters for the binary mixture at a surface to an overall understanding of the interfacial phase behaviour. For the present system, the inputs to the model are the parameters in the pair potentials, $\varepsilon_{AA}$, $\varepsilon_{AB}$, $\varepsilon_{BB}$, $\varepsilon_{wA}$ and $\varepsilon_{wB}$, together with the state point parameters, i.e.\ the temperature $T$ and chemical potentials $\mu_A$ and $\mu_B$ (or, equivalently, the pressure). From these inputs, the present theory yields the interfacial tensions and binding potential. From these (via Young's equation) one can determine other related thermodynamic quantities, such as contact angles, if required. It is through comparing quantities such as contact angles that results from lattice-gas models of the type used here can be related to experiments; see, e.g., Ref.~\cite{perez2021changing}. The interaction parameters can also be estimated from the bulk liquid phase behaviour for both miscible and immiscible liquids; see also the discussion in Ref.~\cite{malijevsky2013sedimentation} about determining these coefficients.

\red{The work of Perez et al.~\cite{perez2021changing} provides a good example of how to select the various interaction parameters in the lattice DFT in order to match results from a specific set of experiments. Our aim here has been different: instead of trying to match any particular experiment, we have sought to explore and illustrate the range of possible behaviours that can be observed as the various interaction strength parameters in the model are changed. This will guide future works towards the right choices to make when applying the model to a specific system. The experimental literature on binary liquids in contact with surfaces is vast. One system that has been well studied is mixtures of lutidine and water on contact with glass \cite{pohl1982wetting} and silica surfaces \cite{beysens1985adsorption}. In the latter experiments, the solid walls under consideration actually correspond to the surfaces of colloidal particles suspended within the liquid mixture. There the interest was focussed on the demixing critical point region of the solvent phase diagram. Similar experiments with different liquid mixtures and particle combinations have also been reported \cite{jayalakshmi1997phase, guo2008reversible, bonn2009direct, nguyen2013controlling} and of particular relevance here, a lattice model closely related to that used here was applied to understand the colloid aggregation behaviour \cite{edison2015critical, edison2015phase, tasios2016critical}. Examples of other binary mixtures where the wetting behaviour has been studied include: nitromethane and carbon disulfide \cite{wu1986wetting, abeysuriya1987control, durian1987wetting}, methanol and cyclohexane \cite{moldover1980interface}, various hydrocarbon and fluorocarbon mixtures \cite{beaglehole1982thickness}, water and propionic acid (and various other mixtures) \cite{nuthalapati2023atypical}, polymer and solvent mixtures \cite{brochard2000wetting} and mixtures of liquid metals \cite{DGGR2021l}.}

\red{The results presented here} show how the density profiles change in form as the attraction between the surface and the two different species of particles in the system varies. As the wall attraction strength increases, so of course does the tendency for the liquid to wet the surface. We also observe that for miscible liquid mixtures, the wall only needs to attract one of the two species strongly for the mixture to wet the surface; it is not necessary to be strongly attractive to both. In contrast, when the wall is weakly attractive, then the liquid(s) do not wet the wall and metastable states where there are bubbles of vapour adsorbed on the surface are possible.

The results presented here for a mixture of immiscible liquids at a surface show that there are multiple possible configurations of the liquids at the surface, and that the interface potential correspondingly has multiple solution branches, which is a nontrivial and rather striking observation. Additionally, we considered cases where the density distributions of the liquids vary in the directions parallel to the surface, which further demonstrates the varied number of possible states, such as droplets on films, droplets on droplets, droplets inside droplets, etc. Our results here point to the free energy landscape for such cases to be immensely complicated with a large number of possible metastable states that deserves further future investigation. \red{As also concluded in Ref.~\cite{luengo2021lifshitz}, the multivalued nature of the binding potentials is likely key to understanding the ubiquity of irreversible wetting phenomena in complex mixtures, where multicomponent solutions frequently exhibit multiphase coexistence.}

Beyond using the binding potentials calculated here to facilitate easy scrutiny of the wetting behaviour of binary liquid mixtures on surfaces, they can also be taken as an input into mesoscopic models like that in Ref.~\cite{ThTL2013prl}, where the focus is on the lowest energy branch of the interface potential. See also the discussion in Ref.~\cite{hughes2015liquid}. However, as future work it would be interesting to examine the full implications of the multi-valued nature of the interface potential when introduced into such mesoscopic models. Such models are also in principle able to describe the nonequilbrium dynamics of liquids droplets on solid substrates; it would be interesting future work to test the reliability of the dynamical predictions of these\red{, for example to shed further light on intriguing recent experiments on mixtures of isopropanol and 2-butanol \cite{pahlavan2021evaporation}, where evaporating droplets are seen to be very far from hemispherical.}

\section*{Acknowledgements}

UT acknowledges support by the Deutsche Forschungsgemeinschaft (DFG, Grant No.\  TH781/12-2 within SPP~2171).


%

\end{document}